\documentclass[aps,prb,twocolumn,superscriptaddress,floatfix,letter]{revtex4-1}
\usepackage{epsfig,amsmath,amssymb,color,dsfont}
\definecolor{jadclr}{rgb}{0,0.5,0}
\definecolor{jadcolor}{rgb}{0.5,0,0}
\usepackage[bookmarks=true,colorlinks,linkcolor=jadclr,urlcolor=jadcolor,citecolor=jadcolor]{hyperref}
\usepackage{cleveref}
\usepackage{bm}
\usepackage{booktabs}
\usepackage{mathrsfs}
\usepackage{array}
\def\d{\mathrm d}

\makeatletter
\newcommand{\thickhline}{%
    \noalign {\ifnum 0=`}\fi \hrule height 1pt
    \futurelet \reserved@a \@xhline
}
\newcolumntype{"}{@{\hskip\tabcolsep\vrule width 1pt\hskip\tabcolsep}}
\makeatother
\begin{document}

\title{Rapid filling of the spin gap with temperature in the Schwinger-boson mean-field theory of the antiferromagnetic Heisenberg kagome model}
\author{Jad C.~Halimeh}
\affiliation{Max Planck Institute for the Physics of Complex Systems, N\"othnitzer Stra\ss e 38, 01187 Dresden, Germany}
\affiliation{Physics Department, Technical University of Munich, 85747 Garching, Germany}

\author{Rajiv R.~P.~Singh}
\affiliation{Department of Physics, University of California Davis, CA 95616, USA}

\date{\today}

\begin{abstract}
Using Schwinger-boson mean-field theory, we calculate the dynamic spin structure factor at low temperatures $0<T\ll J$ for the spin-$1/2$ antiferromagnetic Heisenberg kagome model, within the gapped $\mathbb{Z}_2$ spin liquid phase Ansatz. We find that the spectral gap rapidly fills with temperature, with robust low-energy spectral weight developing by a temperature of $\Delta/3$, where the spin gap is $2\Delta$ (i.e., $\Delta$ is the spinon gap), before any appreciable rise in spinon density or change in zero-temperature mean-field parameters. This is due to deconfinement of spinons which leads to terms suppressed only by $\exp(-\Delta/T)$. At still higher temperatures, the spinon density increases rapidly leading to a breakdown of the Schwinger-boson mean-field approach. We suggest that if the impurity-free spectral functions can be obtained through neutron scattering experiments on kagome herbertsmithites, temperature dependence of the subgap weight can provide distinct signatures of a $\mathbb{Z}_2$ quantum spin liquid.
\end{abstract}

\maketitle
\section{Introduction}\label{sec:Intro}
The Mermin-Wagner theorem\cite{Mermin1966} asserts that in two-dimensional lattices with short-range interactions there can be no spontaneous breaking of continuous symmetries at finite temperatures $T>0$, although such spontaneous symmetry breaking is allowed at $T=0$. 
However, in certain such lattices, \textit{geometric frustration}\cite{Ramirez1994,Mila2000,diep2004,Sachdev2008,lacroix2011,Starykh2015,Balents2010,Savary2017,Zhou2017,Vojta2018} due to the interplay of lattice geometry and antiferromagnetic coupling leads to quantum fluctuations strong enough to preserve continuous symmetries even at $T=0$. A quantum spin liquid\cite{Balents2010,Savary2017,Zhou2017,Vojta2018,Knolle2018} (QSL) is such a phase of matter, where localized magnetic moments are highly correlated but their fluctuations are nevertheless still very pronounced even at $T=0$, leading to a high density of low-lying energy eigenstates, and the ground state can then host fractionalized excitations and topological order.

The ground state of the paradigmatic spin-$1/2$ antiferromagnetic Heisenberg kagome model (AFKM) is a promising candidate for a QSL,\cite{Sachdev1992,Norman2016} while experiments\cite{Shores2005,Helton2007,Mendels2007,Zorko2008,deVries2009,Powell2011,Jeong2011,Zorko2017} on the kagome-lattice compound herbertsmithite indicate that it may indeed comprise such a QSL ground state. A big debate, both experimentally and theoretically, is the existence of a spin gap in the system. 
NMR measurements of Fu \textit{et al.}\cite{Fu2015}~indicate a nonzero spin gap, whereas inelastic neutron scattering (INS) measurements 
of Han \textit{et al.}~suggest a continuum of fractionalized spinon excitations\cite{Han2012} with an absence of any sharp onset with frequency,\cite{Punk2014, Han2016} although it is to be mentioned that INS continua do not necessarily come from fractionalized excitations only. A large number of low-lying excitations can also give a broad frequency response in INS. It is also worth noting here that herbertsmithite is known to be more complex than the nearest-neighbor AFKM primarily due to Dzyaloshinskii-Moriya interactions and impurities,\cite{Dzyaloshinsky1958,Moriya1960,Shekhtman1992,Elhajal2002,Rigol2007,
Messio2010,Huh2010,Dodds2013,Hering2017,Messio2017} and that recent measurements on variants of the herbertsmithite materials
show evidence for gapless excitations.\cite{Gomilsek2017,Orain2017}
On the theoretical side, density matrix renormalization group (DMRG) simulations offer strong evidence for a robustly gapped $\mathbb{Z}_2$ QSL,\cite{Yan2011,Jiang2012,Depenbrock2012,Kolley2015} while many recent computational studies have argued for a gapless, possibly $U(1)$ Dirac QSL state. \cite{Ran2007,Iqbal2013,He2017,Liao2017}

The two-dimensional (2D) $\mathbb{Z}_2$ QSL is known not to need to go through a transition as the temperature is increased, because the involved topological defects are piontlike objects known as visons, which are always created with finite density at nonzero temperatures.\cite{Savary2017,Savary2013} This means that the 2D $\mathbb{Z}_2$ QSL may be smoothly connected to a trivial paramagnet, i.e.~there is only a crossover at finite temperatures. Starting from the ground state with gapped spinon and vison excitations as is the case in a gapped $\mathbb{Z}_2$ QSL, as the temperature is subsequently cranked up, these excitations become thermally populated. As soon as there is a density of thermally excited visons, the different topological ground-state sectors can no longer be distinguished. However, even though strictly speaking the topological order of the 2D gapped $\mathbb{Z}_2$ QSL is destroyed at any finite temperature,\cite{Castelnovo2007,Hastings2011} remnants of the QSL phase must survive in the form of local physical observables, which cannot be immediately destroyed at $T>0$ in the absence of a zero-temperature phase transition.

The dynamic spin structure factor (DSF) offers a useful way of relating theoretical results to INS measurements that can shed light on the properties of the AFKM, and has been numerically computed in this model at zero and finite temperatures using exact diagonalization methods in small systems,\cite{Laeuchli2009,Seman2015,Shimokawa2016} and at zero temperature using Abrikosov fermion mean-field theory\cite{Dodds2013} and SBMFT.\cite{Punk2014,Halimeh2016,Messio2017} Such finite-temperature measurements can allow for a better characterization of the ground-state properties of the AFKM in light of the aforementioned discussion of how QSL behavior at finite temperature is related to the zero-temperature physics. Recently, Ref.~\onlinecite{Sherman2018} has computed the finite-temperature DSF of the AFKM at finite temperatures using the numerical linked cluster expansion (NLCE) method, but the latter is only valid for $T\geq J/4$. Previously, the finite-temperature static structure factor was computed using a high-temperature expansion.\cite{Elstner1994}

In this paper, we compute the finite-temperature DSF of the AFKM in the framework of Schwinger-boson mean-field theory\cite{Arovas1988,Sachdev1992,Auerbach1994,Vishwanath2006,Auerbach2011} (SBMFT). Low-temperature thermodynamic properties have previously been computed in SBMFT, such as in the case of the triangular-lattice \cite{Manuel1998,Mezio2012} and square-lattice\cite{Auerbach1988,Auerbach1994} Heisenberg antiferromagnets. Our work is fundamentally different though, as the latter studies investigate systems that are ordered, and hence gapless in an SBMFT sense, at zero temperature, which leads to subtleties in the SBMFT treatment since at finite temperature a gap suddenly emerges due to being in a disordered phase. On the other hand, AFKM is still gapped and in a disordered phase at $T=0$, and thus we do not face such issues. The work presented here follows zero-temperature DSF calculations\cite{Halimeh2016} in SBMFT of various ground states of the AFKM based on two prototypical An\"atze\cite{Sachdev1992,Messio2013} of the projective symmetry group\cite{Wen2002,Vishwanath2006} (PSG). We are not aware of any previous such calculation at very low but nonzero temperatures. 

In the framework of SBMFT, bond mean fields are used to characterize the QSL, where a given PSG Ansatz sets the properties of the mean fields. These local observables are expected to not vanish immediately at finite temperature due to the crossover from a QSL ground state to a trivial paramagnet. Thus, so long as the spinon density is low enough such that interactions can be neglected, SBMFT can provide a suitable method to qualitatively study AFKM properties at low temperatures.

\subsection{Summary of results}
Our most surprising and striking result is that the spectral-weight in the spin gap ($\sim2\Delta$, where $\Delta$ is the spinon gap) in the DSF fills up rapidly with temperature. Well below the spin-gap energy and even before the SBMFT parameters have changed significantly from their $T=0$ values or there is any significant rise in spinon density, the low-frequency spectral weight starts to get populated. This is due to deconfinement of spinons in a $\mathbb{Z}_2$ QSL, which leads to terms suppressed by a factor of only $\exp(-\Delta/T)$ rather than $\exp(-2\Delta/T)$, the suppression factor in case of confined spinons. Only at still higher temperatures ($T>0.1J$) does the spinon density start rising rapidly leading to a breakdown of the SBMFT treatment. This result applies to both the different mean-field Ans\"atze that we consider. We also note some interesting changes in spectral weight with frequency and wave vector in the Brillouin zone. 

A quantitative comparison of our results with experiments is not appropriate as experimental systems have many additional interactions and also because the SBMFT is not expected to be quantitatively accurate for the spin-half model.
However, the fact that the spin gap is rapidly populated at low temperatures, with an activation energy different from the $T=0$ spin gap, in itself constitutes a signature of deconfinement. This is a robust result and can, in principle, be looked for in experiments. 
However, this is not possible for current experiments in Ref.~\onlinecite{Han2012} where impurities need to be subtracted \cite{Norman2016}
and the very existence of a spin gap is unclear. But, we can still attempt a qualitative comparison. 
As we show below for one of the Ans\"atze, we can qualitatively capture their DSF measurement at low temperature $T\sim J/100$ and low frequency $\omega\sim J/10$. However, our DSF is not constant over frequency as theirs is, but we argue that this can be reproduced in SBMFT by allowing for spinon-vison interactions as is done in Ref.~\onlinecite{Punk2014}. In the latter, the DSF is structureless and flattens at intermediate energies upon including the spinon-vison interactions, albeit there remains an onset around $\omega\sim J/10$. Our results, in which the onset completely vanishes at low temperatures, strongly indicate that such a study at finite temperature incorporating spinon-vison interactions may lead to a much more complete agreement with the measurements of Ref.~\onlinecite{Han2012}, and we leave this open for future work. 
%Moreover, we also explain how our results can be made to stand in even better qualitative agreement with the experimental results of Ref.~\onlinecite{Han2012} by simply decreasing the SBMFT self-consistent value of the spinon gap, which is known to be an overestimate, without having to include any spinon-vison interactions.

\subsection{Structure of the paper}
The rest of the paper is organized as follows. In Sec.~\ref{sec:method}, after introducing the AFKM, we provide a brief review of SBMFT, derive the mean field-decoupled AFKM Hamiltonian, and discuss the self-consistency conditions on the respective bond mean fields and local constraint. In Sec.~\ref{sec:ssf} we derive the finite-temperature DSF. Sec.~\ref{sec:results} provides the numerical results of the finite-temperature DSF for two prominent PSG Ans\"atze, followed by a discussion of all the results. We conclude and provide outlook for follow-up work in Sec.~\ref{sec:conclusion}. The paper contains four Appendices supplementing the material presented in the main text with further details and results. Furthermore, we set Planck's reduced constant $\hbar$ and Boltzmann's constant $k_\text{B}$ to unity throughout the entire paper.

\begin{figure}[t]
 \centering
 \includegraphics[width=0.85\columnwidth]{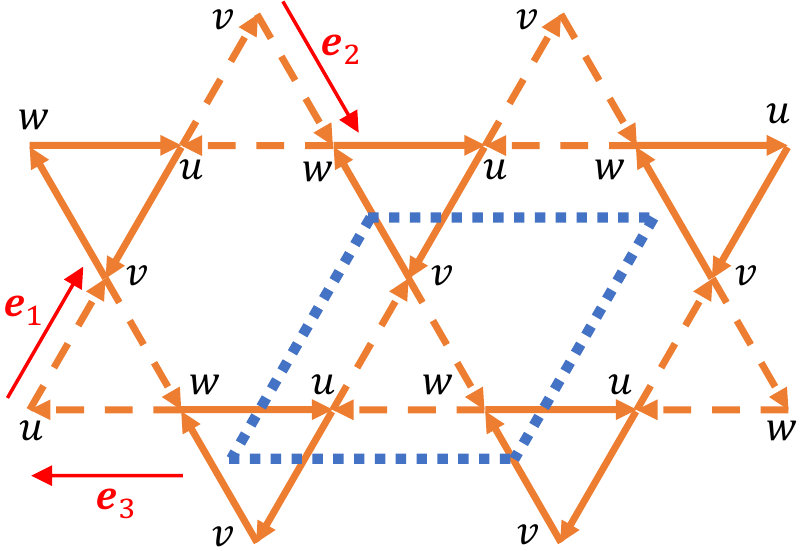}
 \caption{(Color online). The Ans\"atze $\mathbf{q}=\bm{0}$ and $\sqrt{3}\times\sqrt{3}$ have three-site unit cells (demarcated in dashed blue lines), on the kagome lattice, with each containing six bonds, where each bond has a singlet pairing and hopping mean field. For the $\mathbf{q}=\bm{0}$ Ansatz, all pairing and hopping mean fields equal $\mathcal{A}$ and $\mathcal{B}$, respectively. For the $\sqrt{3}\times\sqrt{3}$ Ansatz, bonds with a dashed (solid) arrow have pairing mean field $\pm\mathcal{A}$ and hopping mean field $\mathcal{B}$.
 }
 \label{fig:unitcell}
\end{figure}
 
\section{Model and methods}\label{sec:method}
The antiferromagnetic Heisenberg Hamiltonian on the kagome lattice is given by

\begin{align}\label{eq:H}
\hat{H}=J\sum_{\langle i,j\rangle}\hat{\mathbf{S}}_i\cdot\hat{\mathbf{S}}_j,
\end{align}
where $\hat{\mathbf{S}}_i$ is the spin operator on site $i$, and $J>0$ is the antiferromagnetic spin coupling constant. We now express the spin operators in terms of Schwinger bosons:

\begin{align}
\hat{\mathbf{S}}_i=\frac{1}{2}\hat{b}_{i,\alpha}^\dagger\hat{\bm{\sigma}}^{\alpha\beta}\hat{b}_{i,\beta},
\end{align}
where $\hat{b}_{i,\beta}$ and $\hat{b}_{i,\alpha}^\dagger$ are bosonic annihilation and creation operators satisfying the canonical commutation relations $[\hat{b}_{i,\alpha},\hat{b}_{j,\beta}]=0$ and $[\hat{b}_{i,\alpha},\hat{b}_{j,\beta}^\dagger]=\delta_{i,j}\delta_{\alpha,\beta}$. All throughout the paper, we assume summation over Greek indices, with which we denote the spin degrees of freedom. As such,~\eqref{eq:H} can now be rewritten as

\begin{align}\nonumber
\hat{H}=&\,\frac{J}{4}\sum_{\langle i,j\rangle}\left(2\delta_{\alpha,\mu}\delta_{\beta,\gamma}-\delta_{\alpha,\beta}\delta_{\gamma,\mu}\right)\hat{b}_{i,\alpha}^\dagger\hat{b}_{j,\gamma}^\dagger\hat{b}_{i,\beta}\hat{b}_{j,\mu}\\\label{eq:H_SB}
&+\lambda\sum_i(\hat{b}_{i,\alpha}^\dagger\hat{b}_{i,\alpha}-2\mathcal{S}),
\end{align}
where $\lambda$ is a Lagrange multiplier that constrains, on average, the number of bosons to $2\mathcal{S}$ per site, where $\mathcal{S}$ is the spin length.\cite{Auerbach1988} Note that this is necessary since the Hilbert space of the Schwinger bosons is infinite while that of the spin operators is not. The Lagrange multiplier is a way to make the mapping from spins to Schwinger bosons faithful. Mapping spins to Schwinger bosons has been extensively used in the study of antiferromagnets,\cite{Auerbach1988,Arovas1988,Sachdev1992,Auerbach1994} and has recently also been used in Keldysh quantum field theoretical treatments of out-of-equilibrium strongly-correlated spin systems.\cite{Halimeh2018,Schuckert2018}

Strictly speaking, the Schwinger boson number constraint,
	\begin{align}\label{eq:constraint} \hat{b}_{i,\uparrow}^\dagger\hat{b}_{i,\uparrow}+\hat{b}_{i,\downarrow}^\dagger\hat{b}_{i,\downarrow}=2\mathcal{S},
	\end{align} should be enforced by a site-dependent Lagrange multiplier in~\eqref{eq:H_SB} to enforce exactly $2\mathcal{S}$ bosons per site, but this is numerically very expensive, which is why the site-dependence of $\lambda$ is dropped to enforce this constraint only on average. It is also important to realize that the constraint~\eqref{eq:constraint}, in relating a boson number to a spin length, means that $\mathcal{S}$ can now be treated as a continuous parameter that interpolates between the extreme quantum limit of $\mathcal{S}=0$ and the classical limit of $\mathcal{S}\to\infty$. In the SBMFT treatment of AFKM, choosing the spin length $\mathcal{S}=1/2$ can lead to magnetically ordered phases.\cite{Sachdev1992} It is therefore quite common to go to lower values of $\mathcal{S}$ in order to ensure falling in the QSL phase of this model. For this purpose and for continuity with previous work,\cite{Halimeh2016} in this paper we choose $\mathcal{S}=0.2$, though we stress that other values of $\mathcal{S}<1/2$ can only quantitatively, but not qualitatively, change the main conclusions of this work.

\subsection{Schwinger-boson mean-field theory}
Let us consider the $SU(2)$-symmetric singlet pairing and hopping bond operators

\begin{align}
\hat{\mathcal{A}}_{ij}&=\frac{1}{2}\varepsilon^{\alpha\beta}\hat{b}_{i,\alpha}\hat{b}_{j,\beta},\\
\hat{\mathcal{B}}_{ij}&=\frac{1}{2}\hat{b}_{i,\alpha}^\dagger\hat{b}_{j,\alpha},
\end{align}
respectively, with $\varepsilon^{\alpha\beta}$ the $SU(2)$ Levi-Civita tensor, which allows us to rewrite~\eqref{eq:H_SB} in the form

\begin{align}\label{eq:H_SB2}
\hat{H}=J\sum_{\langle i,j\rangle}(\hat{\mathcal{B}}_{ij}^\dagger\hat{\mathcal{B}}_{ij}-\hat{\mathcal{A}}_{ij}^\dagger\hat{\mathcal{A}}_{ij})+\lambda\sum_i(\hat{b}_{i,\alpha}^\dagger\hat{b}_{i,\alpha}-2\mathcal{S}).
\end{align}
A mean-field decoupling of~\eqref{eq:H_SB2} yields

\begin{align}\nonumber
\hat{H}_\text{MF}=&\,J\sum_{\langle i,j\rangle}(\langle\hat{\mathcal{B}}_{ij}\rangle\hat{\mathcal{B}}_{ij}^\dagger-\langle\hat{\mathcal{A}}_{ij}\rangle\hat{\mathcal{A}}_{ij}^\dagger+\text{H.c.})\\\nonumber
&+J\sum_{\langle i,j\rangle}(\langle\hat{\mathcal{A}}_{ij}\rangle\langle\hat{\mathcal{A}}_{ij}^\dagger\rangle-\langle\hat{\mathcal{B}}_{ij}\rangle\langle\hat{\mathcal{B}}_{ij}^\dagger\rangle)\\\label{eq:H_SBMFT}
&+\lambda\sum_i(\hat{b}_{i,\alpha}^\dagger\hat{b}_{i,\alpha}-2\mathcal{S}).
\end{align}
The fields $\langle\hat{\mathcal{A}}_{ij}\rangle$ and $\langle\hat{\mathcal{B}}_{ij}\rangle$ are in general complex-valued parameters that shall be self-consistently computed at a given temperature $T$ (see Sec.~\ref{sec:SC}). Even though most SBMFT studies usually use only the pairing field $\langle\hat{\mathcal{A}}_{ij}\rangle$, additionally including the hopping field $\langle\hat{\mathcal{B}}_{ij}\rangle$ has been proven\cite{Flint2009,Mezio2011} to offer a better description of the excitation spectrum in frustrated systems.  Fig.~\ref{fig:unitcell} shows the kagome lattice, where a unit cell contains three sites, and thus six bonds. SBMFT involves setting all the auxiliary fields in~\eqref{eq:H_SBMFT} to static and uniform saddle-point (self-consistent) parameters. Here, we consider the two prototypical symmetric Ans\"atze\cite{Sachdev1992,Messio2012,Messio2013} $\mathbf{q}=\bm{0}$ and $\sqrt{3}\times\sqrt{3}$, which are characterized by

\begin{align}
\langle\hat{\mathcal{A}}_{ij}\rangle&=\mathcal{A}\text{e}^{\text{i}\theta},\;\;\;\;\;\langle\hat{\mathcal{B}}_{ij}\rangle=\mathcal{B},
\end{align}
where, consulting Fig.~\ref{fig:unitcell}, $\theta=0$ ($\phi$) on dashed (solid) bonds, with $\phi=0$ for the $\mathbf{q}=\bm{0}$ Ansatz and $\phi=\pi$ for the $\sqrt{3}\times\sqrt{3}$ Ansatz, and the mean fields are such that $\mathcal{A}>0$ and $\mathcal{B}<0$. $\mathcal{A}$, $\mathcal{B}$, and $\lambda$ will be calculated self-consistently for each Ansatz, and their value will depend, in addition to the Ansatz itself, on the temperature $T$ at which our system is. Enforcing self-consistency is discussed in Sec.~\ref{sec:SC}.

We now employ the Fourier transformation

\begin{align}\label{eq:FT}
\hat{b}_{i,\alpha}=\hat{b}_{m,\alpha}^s=\frac{1}{\sqrt{N}}\sum_\mathbf{k}^\text{B.z.}\hat{b}_{\mathbf{k},\alpha}^s\text{e}^{\text{i}\mathbf{k}\cdot(\mathbf{R}_m+\mathbf{s})},
\end{align}
where $N$ is the number of unit cells, the site position is $\mathbf{r}_i=\mathbf{R}_m+\mathbf{s}$, $\mathbf{R}_m$ is the position of the unit cell $m$ housing the site, $\mathbf{s}$ denotes the position of the site within the unit cell, and B.z.~stands for the first Brillouin zone. Plugging~\eqref{eq:FT} into~\eqref{eq:H_SBMFT}, we derive

\begin{align}\label{eq:H_SBMFT_FT}
\hat{H}_\text{MF}=\sum_\mathbf{k}^\text{B.z.}\hat{\Psi}_\mathbf{k}^\dagger D_\mathbf{k}\hat{\Psi}_\mathbf{k}+6NJ\left(\mathcal{A}^2-\mathcal{B}^2\right)-3N\lambda(1+2\mathcal{S}),
\end{align}
where we have introduced the $SU(2)$ spinor

\begingroup
\renewcommand{\arraystretch}{1.5}
\begin{align}
\hat{\Psi}_\mathbf{k}&=
\begin{pmatrix}
	\hat{b}_{\mathbf{k},\uparrow}^u \\
	\hat{b}_{\mathbf{k},\uparrow}^v \\
	\hat{b}_{\mathbf{k},\uparrow}^w \\
	\hat{b}_{-\mathbf{k},\downarrow}^{u\dagger} \\
	\hat{b}_{-\mathbf{k},\downarrow}^{v\dagger} \\
	\hat{b}_{-\mathbf{k},\downarrow}^{w\dagger}
\end{pmatrix},
\end{align}
\endgroup
with
\begingroup
\renewcommand{\arraystretch}{1.5}
\begin{align}
D_\mathbf{k}&=J
\begin{pmatrix}
	\mathcal{B}R_\mathbf{k} & \text{e}^{\text{i}\frac{\phi}{2}}\mathcal{A}P_{\mathbf{k},\phi} \\
	\text{e}^{-\text{i}\frac{\phi}{2}}\mathcal{A}P_{\mathbf{k},\phi}^\intercal & \mathcal{B}R_\mathbf{k}
\end{pmatrix}+\lambda\mathds{1}_6,\\
R_\mathbf{k}&=
\begin{pmatrix}
	0 & \cos k_1 & \cos k_3 \\
	\cos k_1 & 0 & \cos k_2 \\
	\cos k_3 & \cos k_2 & 0
\end{pmatrix},\\
P_{\mathbf{k},\phi}&=
\begin{pmatrix}
	0 & -\cos\big(k_1-\frac{\phi}{2}\big) & \cos\big(k_3+\frac{\phi}{2}\big) \\
	\cos\big(k_1+\frac{\phi}{2}\big) & 0 & -\cos\big(k_2-\frac{\phi}{2}\big) \\
	-\cos\big(k_3-\frac{\phi}{2}\big) & \cos\big(k_2+\frac{\phi}{2}\big) & 0
\end{pmatrix},
\end{align}
\endgroup
where $\mathds{1}_d$, with $d\in\mathbb{N}$, is the $d\times d$ identity matrix and $\phi=0$ or $\pi$ if the Ansatz is $\mathbf{q}=\bm{0}$ or $\sqrt{3}\times\sqrt{3}$, respectively. Moreover, our notation entails denoting $k_j=\mathbf{k}\cdot\mathbf{e}_j$, $j\in\{1,2,3\}$, with the real-space vectors $\mathbf{e}_1=a(1/2,\sqrt{3}/2)$, $\mathbf{e}_2=a(1/2,-\sqrt{3}/2)$, and $\mathbf{e}_3=a(-1,0)$, and $a$ is the intersite spacing, which, without any loss of generality, we set to unity throughout the paper.

\subsection{Bogoliubov transformation}
We now diagonalize~\eqref{eq:H_SBMFT_FT} by employing the Bogoliubov transformation

\begin{align}\label{eq:Bogo}
\hat{\Psi}_\mathbf{k}=M_\mathbf{k}\hat{\Gamma}_\mathbf{k},
\end{align}
with
\begingroup
\renewcommand{\arraystretch}{1.5}
\begin{align}\label{eq:BogoMatrix}
M_\mathbf{k}=
\begin{pmatrix}
	U_\mathbf{k} & X_\mathbf{k} \\
	V_\mathbf{k} & Y_\mathbf{k}
\end{pmatrix},
\end{align}
\endgroup
and the Bogoliubov spinor

\begingroup
\renewcommand{\arraystretch}{1.5}
\begin{align}
\hat{\Gamma}_\mathbf{k}=
\begin{pmatrix}
	\hat{\gamma}_{\mathbf{k},\uparrow}^u \\
	\hat{\gamma}_{\mathbf{k},\uparrow}^v \\
	\hat{\gamma}_{\mathbf{k},\uparrow}^w \\
	\hat{\gamma}_{-\mathbf{k},\downarrow}^{u\dagger} \\
	\hat{\gamma}_{-\mathbf{k},\downarrow}^{v\dagger} \\
	\hat{\gamma}_{-\mathbf{k},\downarrow}^{w\dagger}
\end{pmatrix},
\end{align}
\endgroup
where the Bogoliubov operators satisfy the canonical commutation relations $[\hat{\gamma}_{\mathbf{k},\alpha},\hat{\gamma}_{\mathbf{q},\beta}]=0$ and $[\hat{\gamma}_{\mathbf{k},\alpha},\hat{\gamma}_{\mathbf{q},\beta}^\dagger]=\delta_{\mathbf{k},\mathbf{q}}\delta_{\alpha,\beta}$. The Bogoliubov transformation~\eqref{eq:Bogo} diagionalizes~\eqref{eq:H_SBMFT_FT} if and only if

\begingroup
\renewcommand{\arraystretch}{1.5}
\begin{align}\label{eq:CondForM}
M_\mathbf{k}^\dagger\tau^3M_\mathbf{k}&=\tau^3,\\
M_\mathbf{k}^\dagger D_\mathbf{k}M_\mathbf{k}&=\mathscr{E}_\mathbf{k}=
\begin{pmatrix}
	\mathcal{E}_{\mathbf{k},\uparrow} & \bm{0}_3 \\
	\bm{0}_3 & \mathcal{E}_{-\mathbf{k},\downarrow}
\end{pmatrix},
\end{align}
\endgroup
where
\begingroup
\renewcommand{\arraystretch}{1.5}
\begin{align}
\tau^3&=
\begin{pmatrix}
	\mathds{1}_3 & \bm{0}_3 \\
	\bm{0}_3 & -\mathds{1}_3
\end{pmatrix},
\end{align}
\endgroup
$\bm{0}_3$ is the $3\times3$ zero matrix, and

\begingroup
\renewcommand{\arraystretch}{1.5}
\begin{align}
\mathcal{E}_{\mathbf{q},\alpha}&=
\begin{pmatrix}
	\epsilon_{\mathbf{q},\alpha}^u & 0 & 0 \\
	0 & \epsilon_{\mathbf{q},\alpha}^v & 0 \\
	0 & 0 & \epsilon_{\mathbf{q},\alpha}^w
\end{pmatrix},
\end{align}
\endgroup
are the Bogoliubov bosonic eigenenergies at momentum $\mathbf{k}$ and spin polarization $\alpha$. We recall here that due to time-reversal invariance and $SU(2)$ symmetry one has $\epsilon_{\mathbf{k},\uparrow}^s=\epsilon_{-\mathbf{k},\downarrow}^s$ and $\epsilon_{\mathbf{k},\uparrow}^s=\epsilon_{\mathbf{k},\downarrow}^s$, respectively, with $s\in\{u,v,w\}$. Even though $M_\mathbf{k}$ can in principle be calculated analytically for both Ans\"atze $\mathbf{q}=\bm{0}$ and $\sqrt{3}\times\sqrt{3}$, it contains very lengthy expressions. Nevertheless, it can be very efficiently and cheaply numerically computed using standard matrix-diagonalization functions in MATLAB or Mathematica, for example. Care has to be taken though so as to ensure that~\eqref{eq:CondForM} is satisfied. Thus, with the Bogoliubov transformation one can rewrite~\eqref{eq:H_SBMFT_FT} in the diagonal form

\begin{align}
\hat{H}_\text{MF}=&\,\sum_\mathbf{k}^\text{B.z.}\hat{\Gamma}_\mathbf{k}^\dagger \mathscr{E}_\mathbf{k}\hat{\Gamma}_\mathbf{k}+6NJ\left(\mathcal{A}^2-\mathcal{B}^2\right)-3N\lambda(1+2\mathcal{S}).
\end{align}
As such, with regards to the time-dependent Bogoliubov operators, we use the Heisenberg equation to derive

\begin{align}\label{eq:HeisSol}
\hat{\gamma}_{\mathbf{k},\alpha}^r(t)=\text{e}^{-\text{i}\epsilon_{\mathbf{k},\alpha}^rt}\hat{\gamma}_{\mathbf{k},\alpha}^r.
\end{align}
This relation will be useful in the derivation of the DSF in Sec.~\ref{sec:ssf}.
\subsection{Self-consistent mean-field parameters}
\label{sec:SC}

\begin{figure}[t]
 \centering
 \includegraphics[width=0.75\columnwidth]{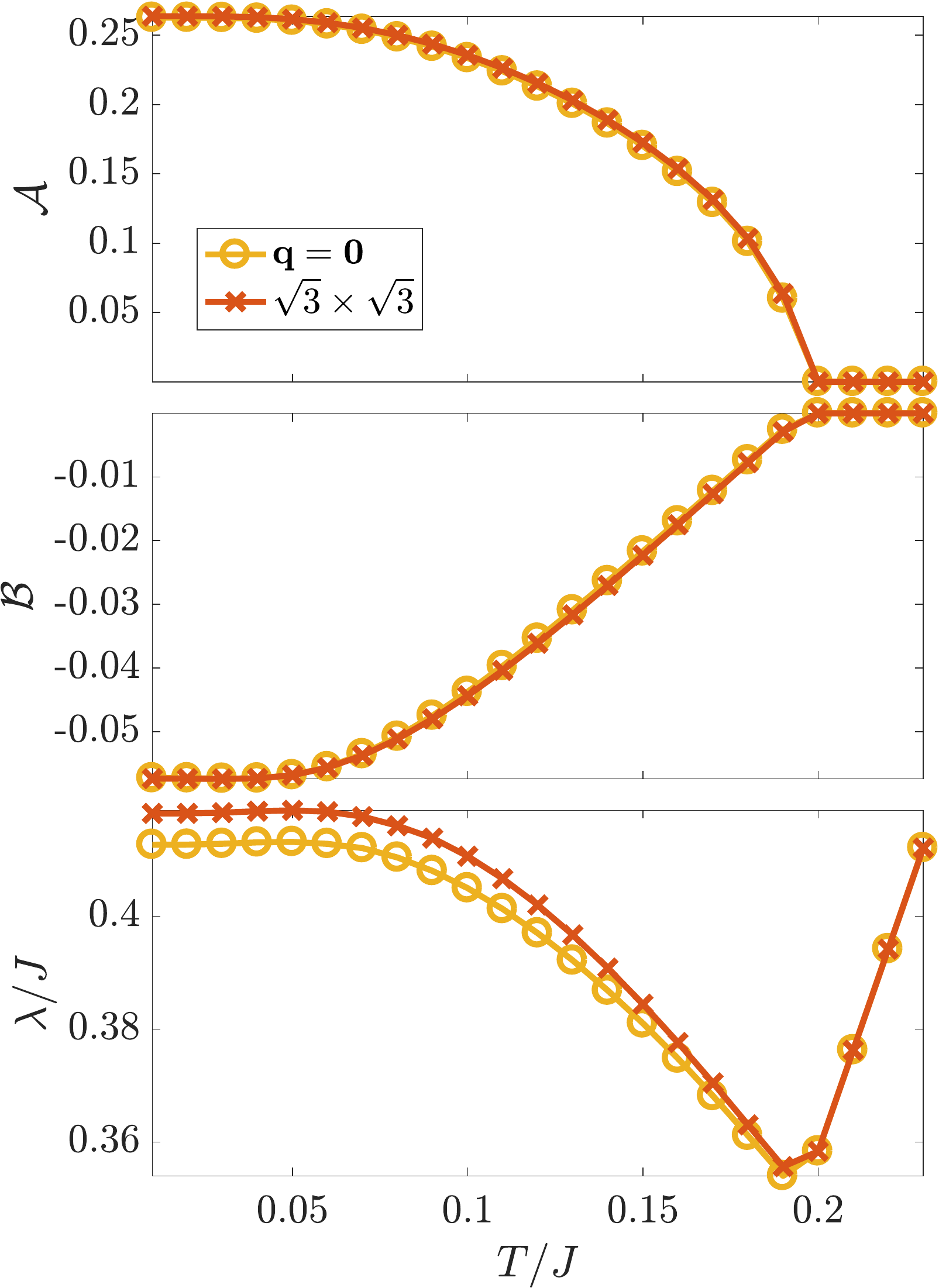}
 \caption{(Color online). Self-consistent bond mean fields and local constraint parameter as function of temperature for the symmetric Ans\"atze $\mathbf{q}=\bm{0}$ and $\sqrt{3}\times\sqrt{3}$. The apparent nonanalyticity at $T\approx J/5$ indicates the unreliability of SBMFT at too high temperatures. In reality, one expects the bond parameters $\mathcal{A}$ and $\mathcal{B}$ to smoothly and asymptotically go to zero as is the case in a crossover.
 }
 \label{fig:fields}
\end{figure}

On a unit cell $m$, the necessary and sufficient conditions for self-consistency for both bond mean fields and the Lagrange multiplier are

\begin{figure}[t]
 \centering
 \includegraphics[width=0.75\columnwidth]{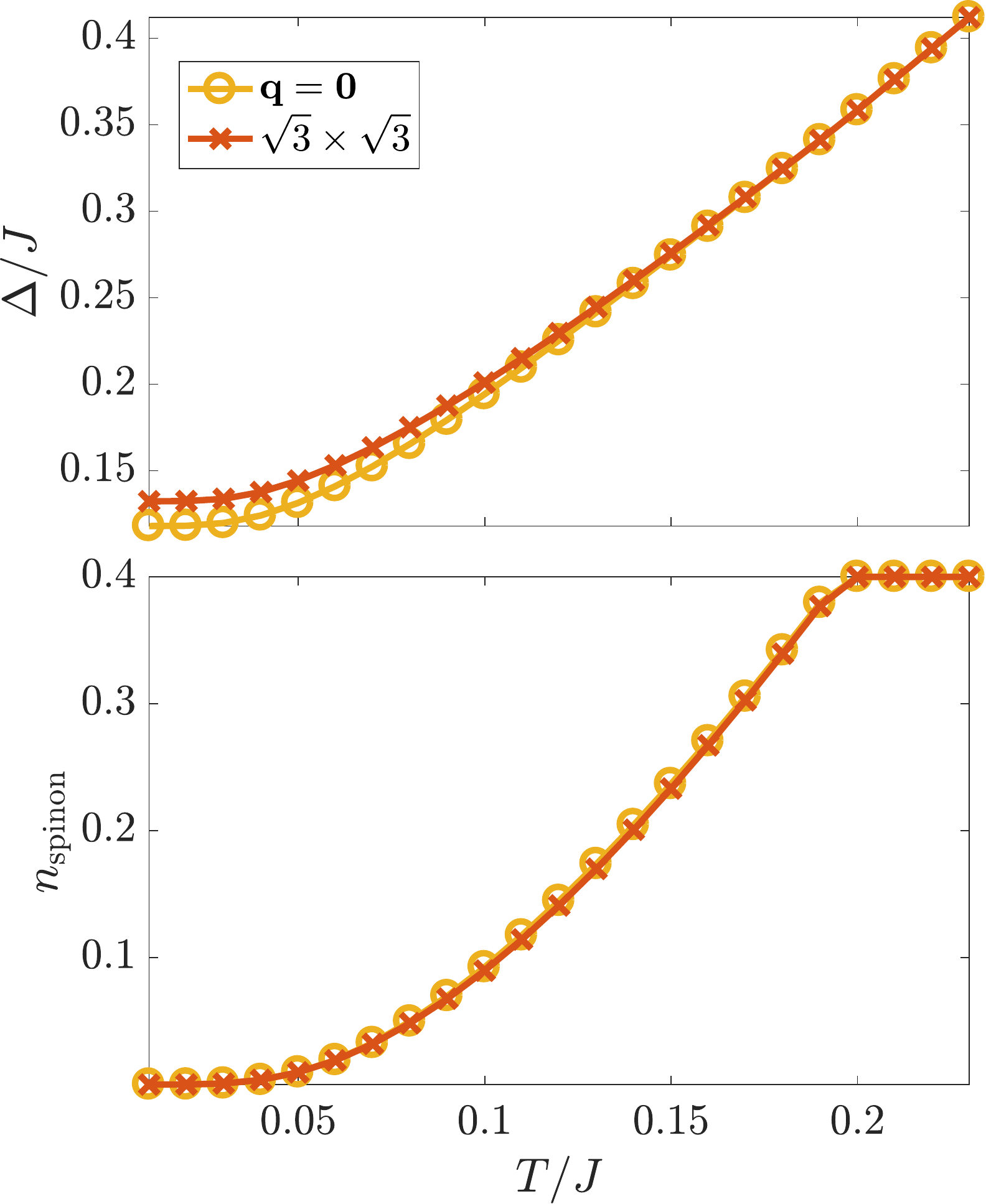}
 \caption{(Color online). The gap (top panel) and spinon density (bottom panel) as function of temperature for the symmetric Ans\"atze $\mathbf{q}=\bm{0}$ and $\sqrt{3}\times\sqrt{3}$ as calculated in SBMFT. The spinon density is small such that interactions can be neglected for low temperatures making SBMFT adequate for the low-temperature description of the AFKM.
 }
 \label{fig:SD}
\end{figure}

\begin{align}\nonumber
\mathcal{A}=&\,\frac{1}{12N}\varepsilon^{\alpha\beta}\sum_m^\text{u.c.}\langle\hat{b}_{m,\alpha}^u\hat{b}_{m,\beta}^v+\hat{b}_{m,\alpha}^v\hat{b}_{m,\beta}^w+\hat{b}_{m,\alpha}^w\hat{b}_{m,\beta}^u\\\label{eq:CondA}
&+\text{e}^{-\text{i}\phi}\left(\hat{b}_{m,\alpha}^u\hat{b}_{\tilde{m},\beta}^v+\hat{b}_{m,\alpha}^v\hat{b}_{\tilde{m},\beta}^w+\hat{b}_{m,\alpha}^w\hat{b}_{\tilde{m},\beta}^u\right)\rangle,\\\nonumber
\mathcal{B}=&\,\frac{1}{12N}\sum_m^\text{u.c.}\langle\hat{b}_{m,\alpha}^{u\dagger}\hat{b}_{m,\alpha}^v+\hat{b}_{m,\alpha}^{v\dagger}\hat{b}_{m,\alpha}^w+\hat{b}_{m,\alpha}^{w\dagger}\hat{b}_{m,\alpha}^u\\\label{eq:CondB}
&+\hat{b}_{m,\alpha}^{u\dagger}\hat{b}_{\tilde{m},\alpha}^v+\hat{b}_{m,\alpha}^{v\dagger}\hat{b}_{\tilde{m},\alpha}^w+\hat{b}_{m,\alpha}^{w\dagger}\hat{b}_{\tilde{m},\alpha}^u\rangle,\\\label{eq:Condlambda}
2\mathcal{S}=&\,\frac{1}{3N}\sum_m^\text{u.c.}\langle\hat{b}_{m,\alpha}^{u\dagger}\hat{b}_{m,\alpha}^u+\hat{b}_{m,\alpha}^{v\dagger}\hat{b}_{m,\alpha}^v+\hat{b}_{m,\alpha}^{w\dagger}\hat{b}_{m,\alpha}^w\rangle,
\end{align}
which are then solved numerically at a given temperature $T$ using fixed-point iteration or some other efficient method. All throughout we assume that a spinon condensate does not form, and this can always be justified so long as we do not get complex spinon eigenvalues. We note that this method is an alternative to the one based on free-energy extremization\cite{Sachdev1992,Messio2013,Halimeh2016} that has traditionally been used, but it gives the same results and is more efficient based on our experience.

We present in Fig.~\ref{fig:fields} the self-consistent field values for spin length $\mathcal{S}=0.2$ and at temperatures up to $T=0.23J$. We see that both bond mean fields $\mathcal{A}$ and $\mathcal{B}$ smoothly decrease in magnitude until $T\approx J/5$ where they nonanalytically go to zero. This is a result of SBMFT being inadequate for the description of the paramagnetic phase at temperatures so high that nearest-neighbor correlations are destroyed.\cite{Auerbach1994} Moreover, it is clear that the bond fields going to zero cannot be an indication of a continuous phase transition for two main reasons: (i) the 2D gapped $\mathbb{Z}_2$ spin liquid does not undergo such a transition, but rather a crossover, to a trivial paramagnet at finite temperature; and (ii) the bond fields $\mathcal{A}$ and $\mathcal{B}$ are not local order parameters in the Landau sense. Despite this nonanalyticity being an artifact of SBMFT at too high temperatures,\cite{Auerbach1994,Arovas1988,Mezio2012} it is known that at low temperatures where the mean fields are nonzero SBMFT gives qualitatively reliable results.\cite{Auerbach1994,Manuel1998,Arovas1988,Mezio2012,Auerbach1988} In fact, Fig.~\ref{fig:SD} shows the spinon gap $\Delta$ and spinon density

\begin{align}
n_\text{spinon}=\frac{1}{3N}\sum_r^\text{bands}\sum_\mathbf{k}^\text{B.z.}  \frac{1}{\text{e}^{\epsilon_{\mathbf{k},\alpha}^r/T}-1},
\end{align}
where it can be seen that for the low temperatures we consider ($T\lesssim J/10$) $n_\text{spinon}$ is small enough such that interactions can be neglected, thus rendering SBMFT results valid. However, for higher temperatures, Fig.~\ref{fig:SD} shows that the spinon density can no longer be considered small enough for interactions to be neglected, which means that SBMFT is not to be considered a faithful description of the underlying physics. More drastically, once the bond fields are completely diminished at $T\approx J/5$, which is the case for a high-temperature trivial paramagnet,\cite{Arovas1988,Auerbach1994} the qualitative validity of SBMFT completely fails. Indeed, when $\mathcal{A}=\mathcal{B}=0$, the Hamiltonian~\eqref{eq:H_SBMFT_FT} is diagonal with only $\lambda$ along the diagonal of $D_\mathbf{k}$. Hence, a Bogoliubov transformation is not needed, and the ``spinon" density is just $2\mathcal{S}$ then. Therefore, here it no longer makes sense to speak of spinons, because in this limit the excitations in SBMFT correspond to simply adding or removing a boson on a lattice site, but these excitations are unphysical and have no correspondence in the physical Hilbert space of the original spin model. There are ways of extending the theory to more reliably handle such high temperatures,\cite{Mezio2012} though for low temperatures SBMFT proper gives qualitatively sound results that are often offset by a trivial factor.\cite{Arovas1988,Auerbach1988} As we are interested only in low-temperature DSF calculations, such extensions to SBMFT are outside the scope of our paper.

\begin{figure*}[htp]
\centering
\hspace{-0.15 cm}
\includegraphics[width=.495\textwidth]{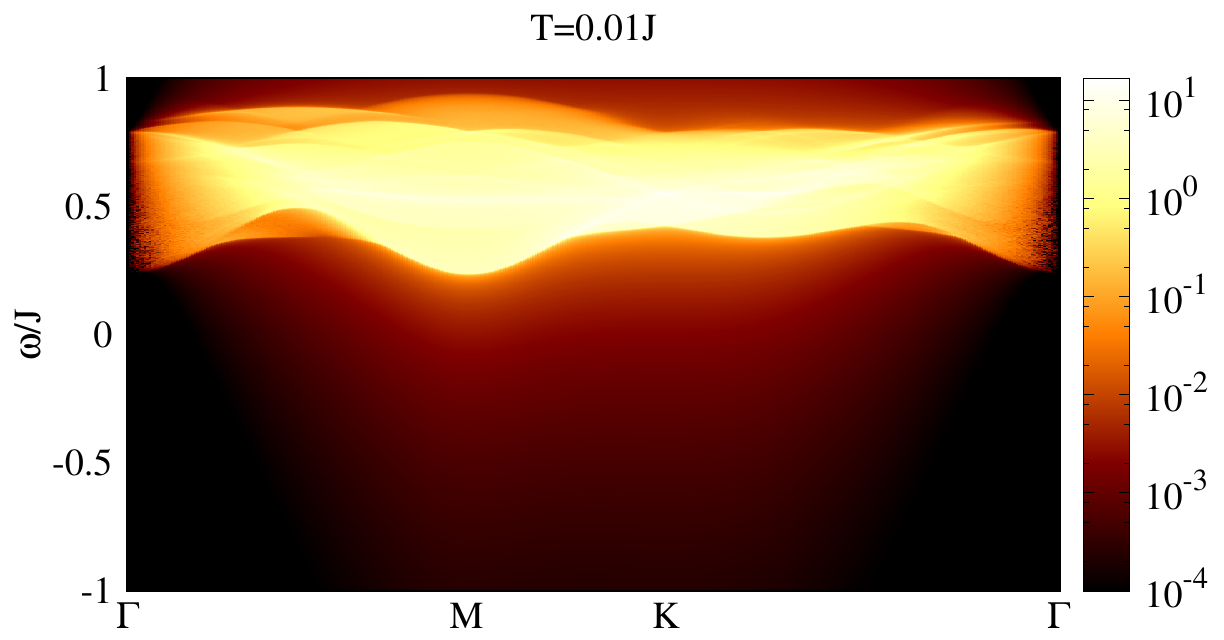}
\hspace{-0.15 cm}
\includegraphics[width=.495\textwidth]{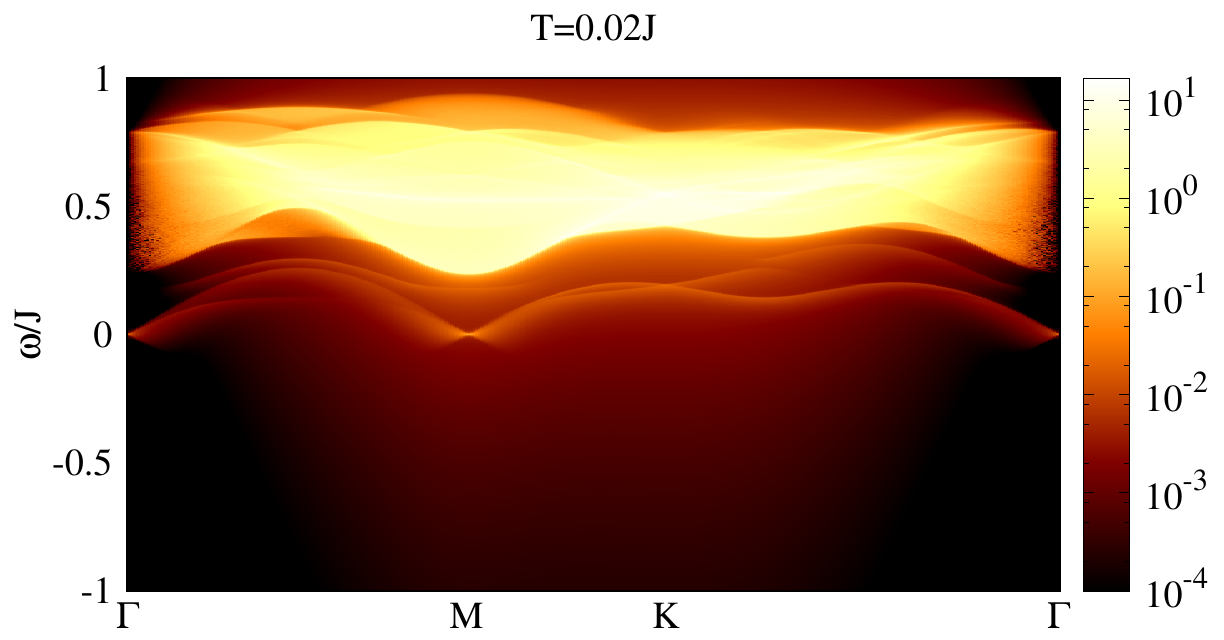}\\
\hspace{-0.15 cm}
\includegraphics[width=.495\textwidth]{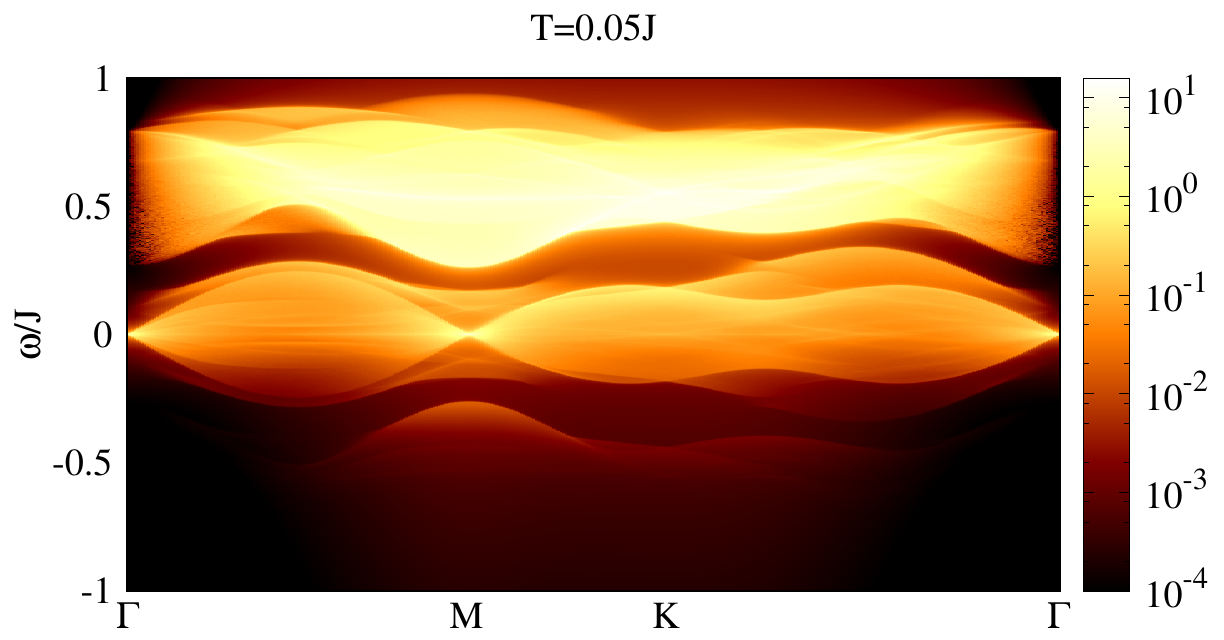}
\hspace{-0.15 cm}
\includegraphics[width=.495\textwidth]{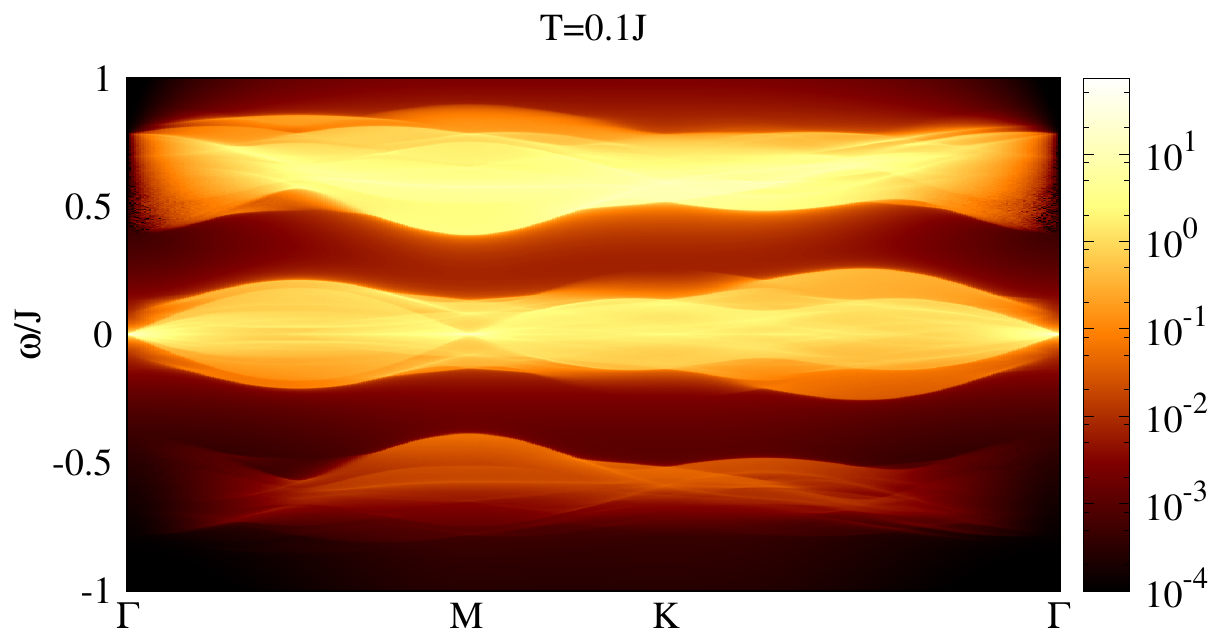}
\caption{(Color online). The dynamic spin structure factor for the $\mathbf{q}=\bm{0}$ Ansatz along the $\Gamma$-M-K-$\Gamma$ high-symmetry lines at finite temperatures $T/J=0.01$, $0.02$, $0.05$, and $0.1$. Even though at $T=0J$ the DSF displays no spectral weight at all below the spin gap,\cite{Halimeh2016} at $T=0.01J$ it already shows nonnegligible weight below the spin gap continuously down to negative frequencies, while already at $T=0.02J$ the DSF below the spin gap shows nontrivial weight.}
\label{fig:DSF_q0} 
\end{figure*}\begin{figure*}[htp]
\centering
\hspace{-0.15 cm}
\includegraphics[width=.351\textwidth]{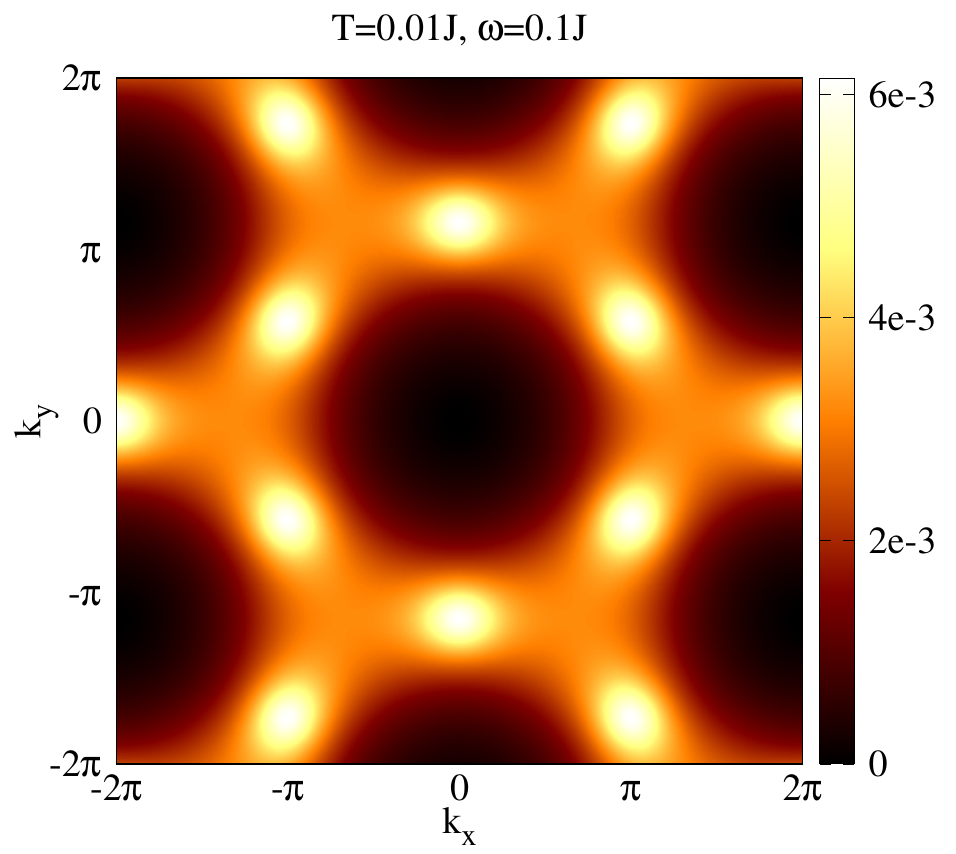}
\hspace{-0.15 cm}
\includegraphics[width=.351\textwidth]{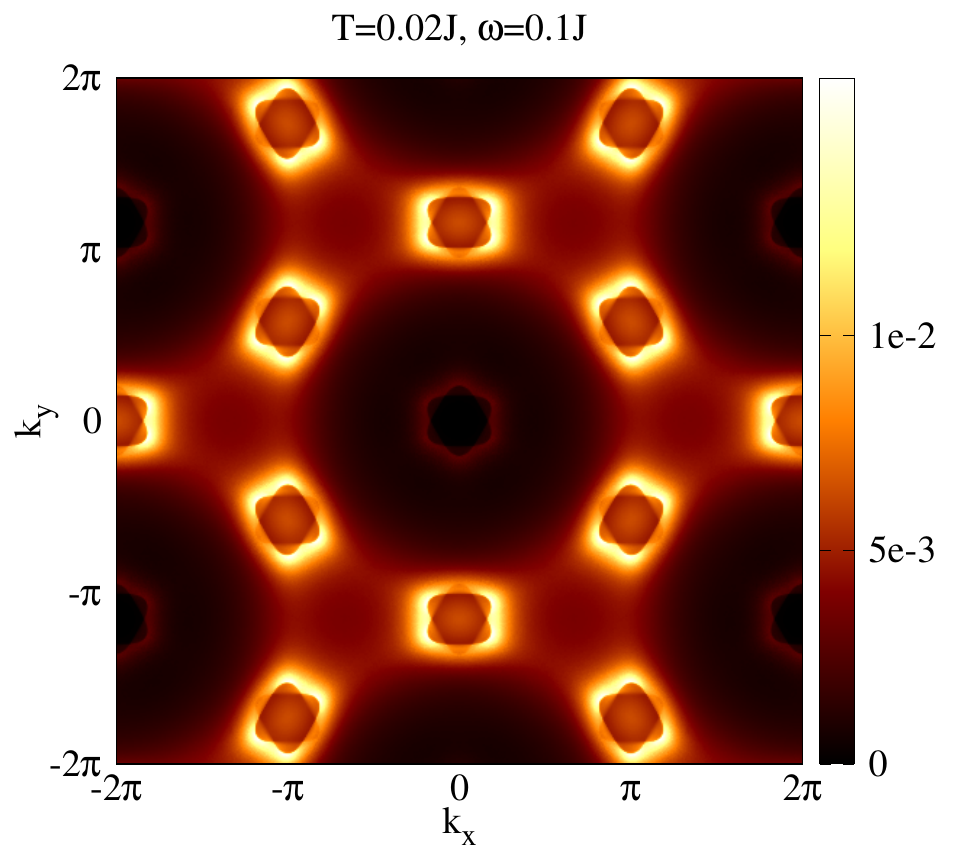}\\
\hspace{-0.15 cm}
\includegraphics[width=.348\textwidth]{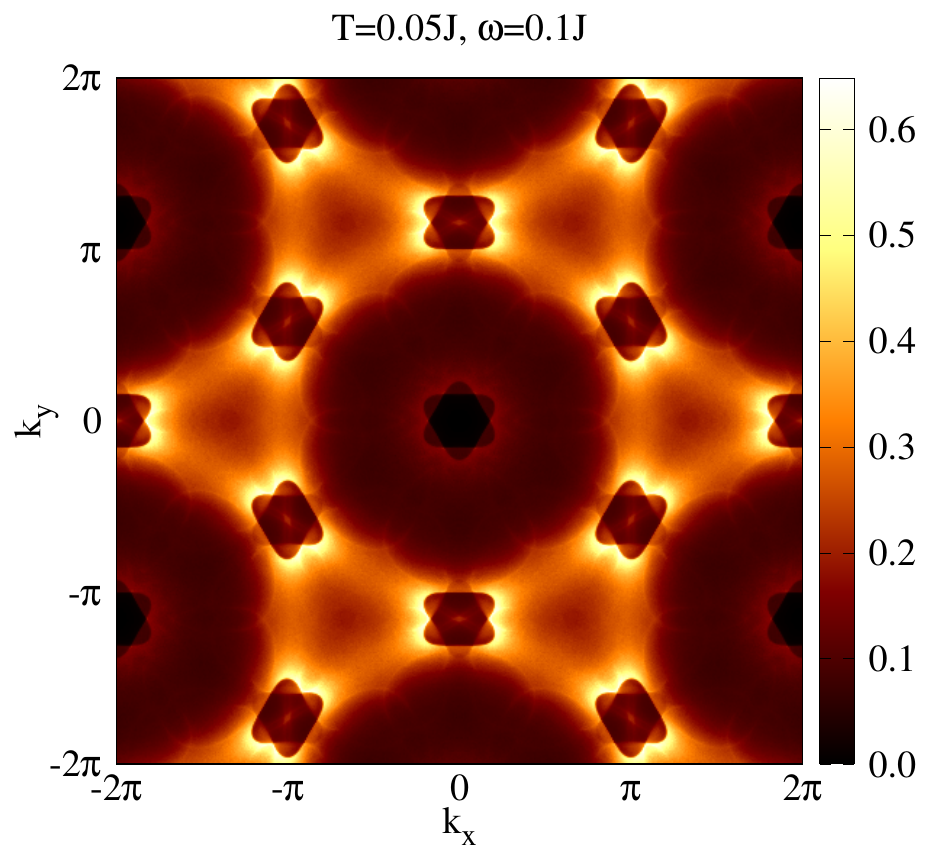}
\hspace{-0.15 cm}
\includegraphics[width=.348\textwidth]{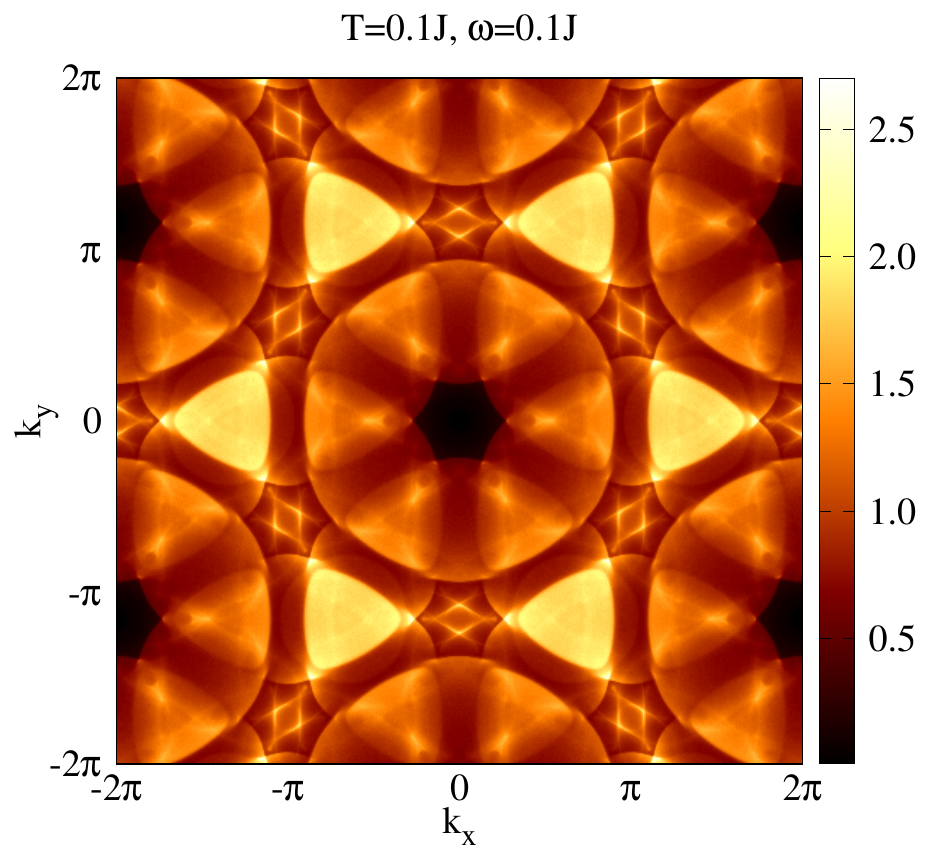}
\caption{(Color online). The dynamic spin structure factor over the extended Brillouin zone at finite temperature and fixed frequency $\omega=0.1J$ for the $\mathbf{q}=\bm{0}$ Ansatz. All results exhibit a six-fold rotation symmetry around the $\Gamma$ point due to time-reversal invariance. Interestingly, the DSF exhibits a rich structure at $T=0.01J$, where it is reminiscent of the experimental result in Fig.~1(c) in Ref.~\onlinecite{Han2012} at an energy roughly an order of magnitude below $J$.}
\label{fig:DSF_q0_omegaLow} 
\end{figure*}

\section{Spin structure factors}\label{sec:ssf}
We now derive the finite-temperature DSF for the AFKM in the framework of SBMFT. The DSF is the Fourier transform of the space-time spin-spin correlations, and is formally given by

\begin{align}
S(\mathbf{k},\omega)=\frac{1}{3N}\sum_{l,j}\text{e}^{-\text{i}\mathbf{k}\cdot(\mathbf{r}_l-\mathbf{r}_j)}\int_{-\infty}^{\infty}\d t\;\text{e}^{\text{i}\omega t} \langle\mathbf{\hat{S}}_l(t)\cdot\mathbf{\hat{S}}_j\rangle.
\end{align}
Recalling that we have $\epsilon_{\mathbf{p},\uparrow}^j=\epsilon_{\mathbf{p},\downarrow}^j$ due to $SU(2)$ symmetry, and employing~\eqref{eq:HeisSol} and the relations

\begin{align}
\langle\hat{\gamma}_{\mathbf{k},\alpha}^{r\dagger}\hat{\gamma}_{\mathbf{q},\beta}^s\rangle&=\frac{1}{\text{e}^{\beta\epsilon_{\mathbf{k},\alpha}^r}-1}\delta_{\mathbf{k},\mathbf{q}}\delta_{r,s}\delta_{\alpha,\beta},\\
\langle\hat{\gamma}_{\mathbf{k},\alpha}^r\hat{\gamma}_{\mathbf{q},\beta}^s\rangle&=0,
\end{align} 
we derive
\begin{widetext}
\begin{align}\nonumber
S(\mathbf{k},\omega)=&\,\frac{1}{12N}\sum_{\mathbf{q}}^\text{B.z.}\sum_{r,s,m,n}^\text{bands}\Bigg[\frac{\delta(\omega+\epsilon_{\mathbf{q},\uparrow}^m+\epsilon_{-\mathbf{k}-\mathbf{q},\uparrow}^n)}{\big(\text{e}^{\epsilon_{\mathbf{q},\uparrow}^m/T}-1\big)\big(\text{e}^{\epsilon_{-\mathbf{k}-\mathbf{q},\uparrow}^n/T}-1\big)}\mathscr{A}^{r,s,m,n}_{\mathbf{k},\mathbf{q}}+\frac{\text{e}^{\epsilon_{\mathbf{k}+\mathbf{q},\uparrow}^n/T}\delta(\omega+\epsilon_{\mathbf{q},\uparrow}^m-\epsilon_{\mathbf{k}+\mathbf{q},\uparrow}^n)}{\big(\text{e}^{\epsilon_{\mathbf{q},\uparrow}^m/T}-1\big)\big(\text{e}^{\epsilon_{\mathbf{k}+\mathbf{q},\uparrow}^n/T}-1\big)}\mathscr{B}^{r,s,m,n}_{\mathbf{k},\mathbf{q}}\\\label{eq:DSF}
%%%%%%%%%%%%%%%%%%%%%%%
&+\frac{\text{e}^{\epsilon_{-\mathbf{q},\uparrow}^m/T}\delta(\omega-\epsilon_{-\mathbf{q},\uparrow}^m+\epsilon_{-\mathbf{k}-\mathbf{q},\uparrow}^n)}{\big(\text{e}^{\beta\epsilon_{-\mathbf{q},\uparrow}^m/T}-1\big)\big(\text{e}^{\epsilon_{-\mathbf{k}-\mathbf{q},\uparrow}^n/T}-1\big)}\mathscr{C}^{r,s,m,n}_{\mathbf{k},\mathbf{q}}+\frac{\text{e}^{\epsilon_{-\mathbf{q},\uparrow}^m/T}\text{e}^{\epsilon_{\mathbf{k}+\mathbf{q},\uparrow}^n/T}\delta(\omega-\epsilon_{-\mathbf{q},\uparrow}^m-\epsilon_{\mathbf{k}+\mathbf{q},\uparrow}^n)}{\big(\text{e}^{\epsilon_{-\mathbf{q},\uparrow}^m/T}-1\big)\big(\text{e}^{\epsilon_{\mathbf{k}+\mathbf{q},\uparrow}^n/T}-1\big)}\mathscr{D}^{r,s,m,n}_{\mathbf{k},\mathbf{q}}\Bigg],
\end{align}
\end{widetext}
where the terms in script font are defined in Appendix~\ref{sec:DSFterms}, and they comprise sums of products of the momentum-dependent Bogoliubov matrices of~\eqref{eq:BogoMatrix}.

\begin{figure*}[htp]
\centering
\hspace{-0.15 cm}
\includegraphics[width=.495\textwidth]{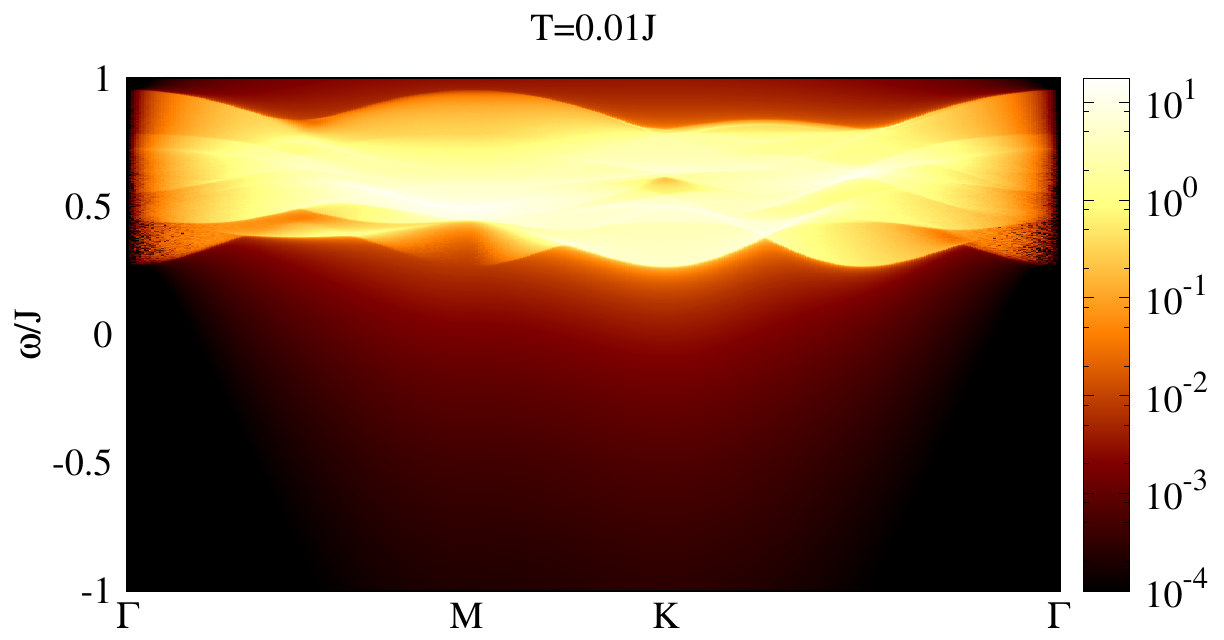}
\hspace{-0.15 cm}
\includegraphics[width=.495\textwidth]{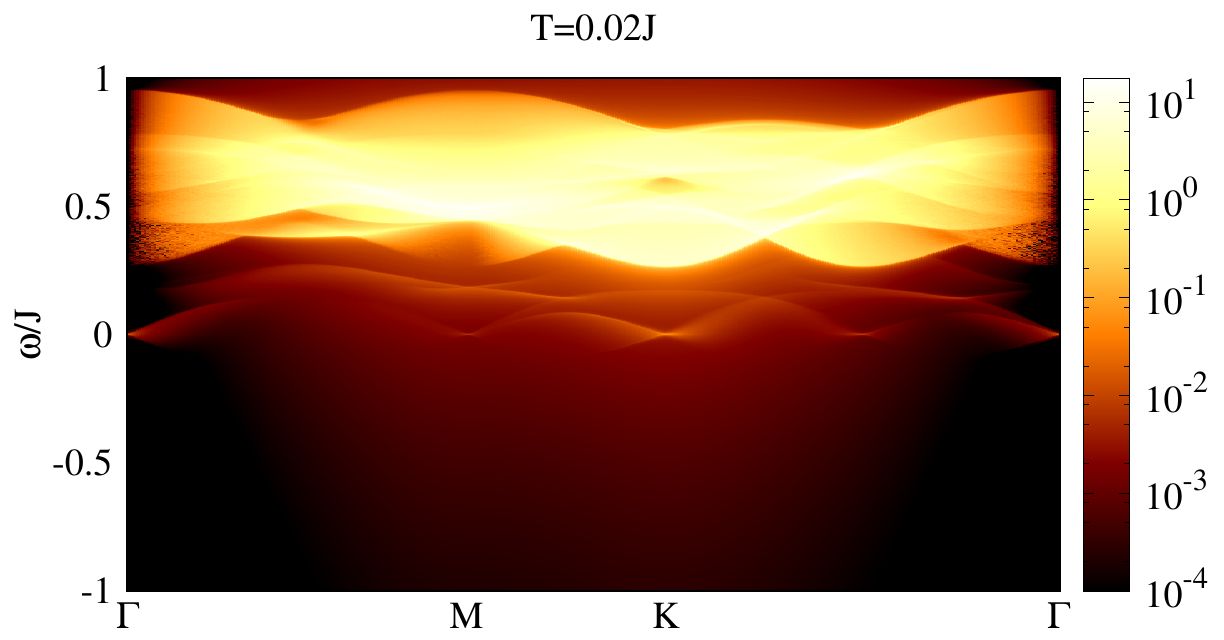}\\
\hspace{-0.15 cm}
\includegraphics[width=.495\textwidth]{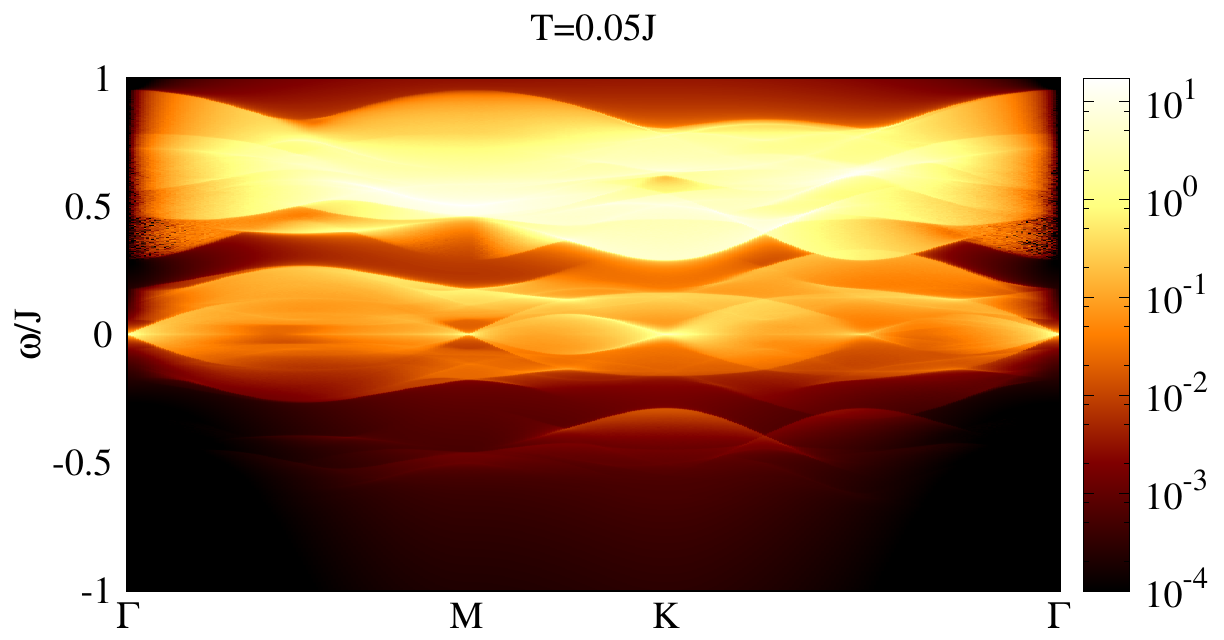}
\hspace{-0.15 cm}
\includegraphics[width=.495\textwidth]{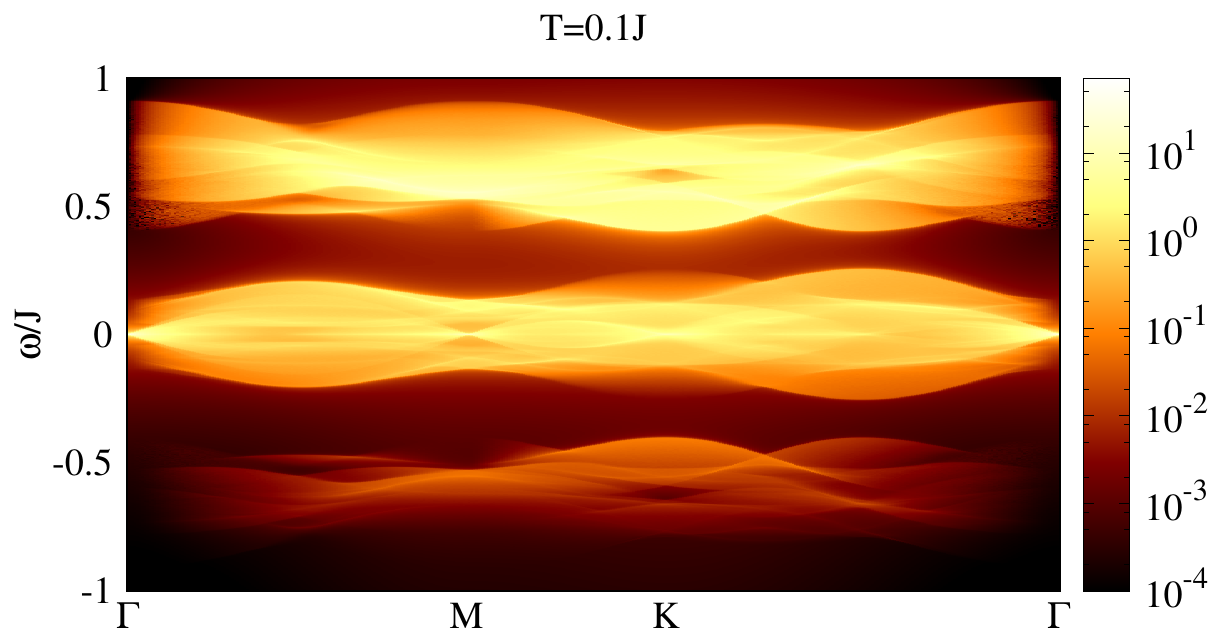}
\caption{(Color online). Same as Fig.~\ref{fig:DSF_q0} but for the $\sqrt{3}\times\sqrt{3}$ Ansatz.}
\label{fig:DSF_Sqrt3} 
\end{figure*}

\begin{figure*}[htp]
\centering
\includegraphics[width=.351\textwidth]{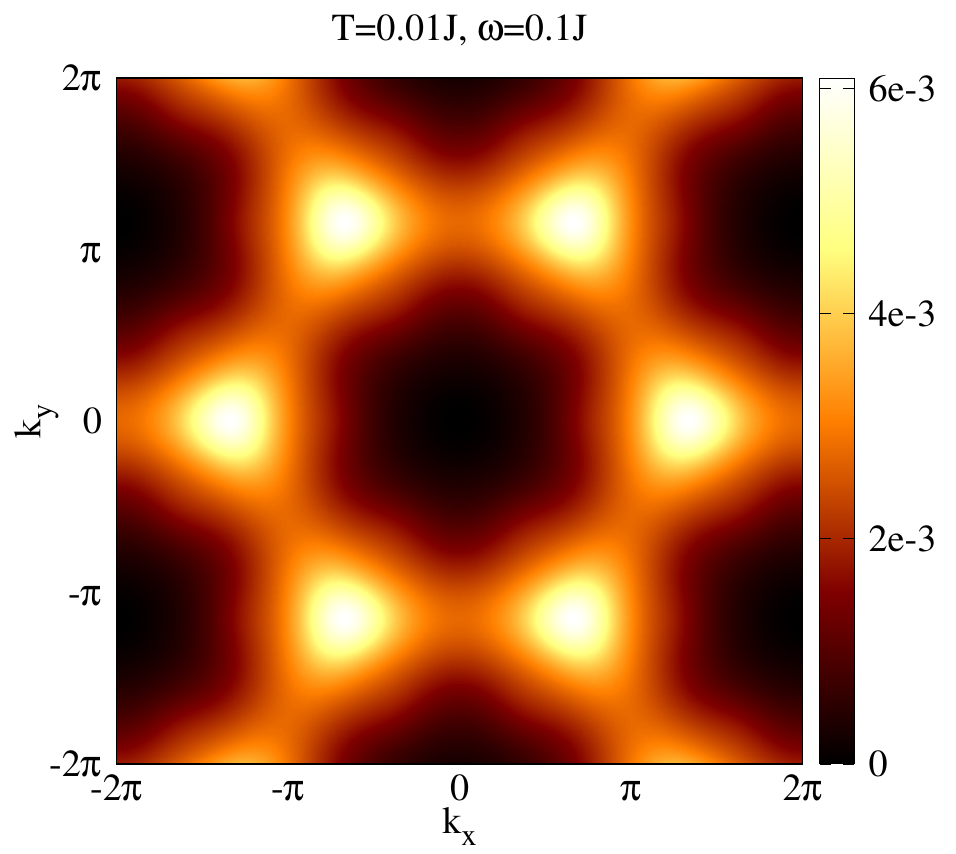}
\hspace{-0.15 cm}
\includegraphics[width=.351\textwidth]{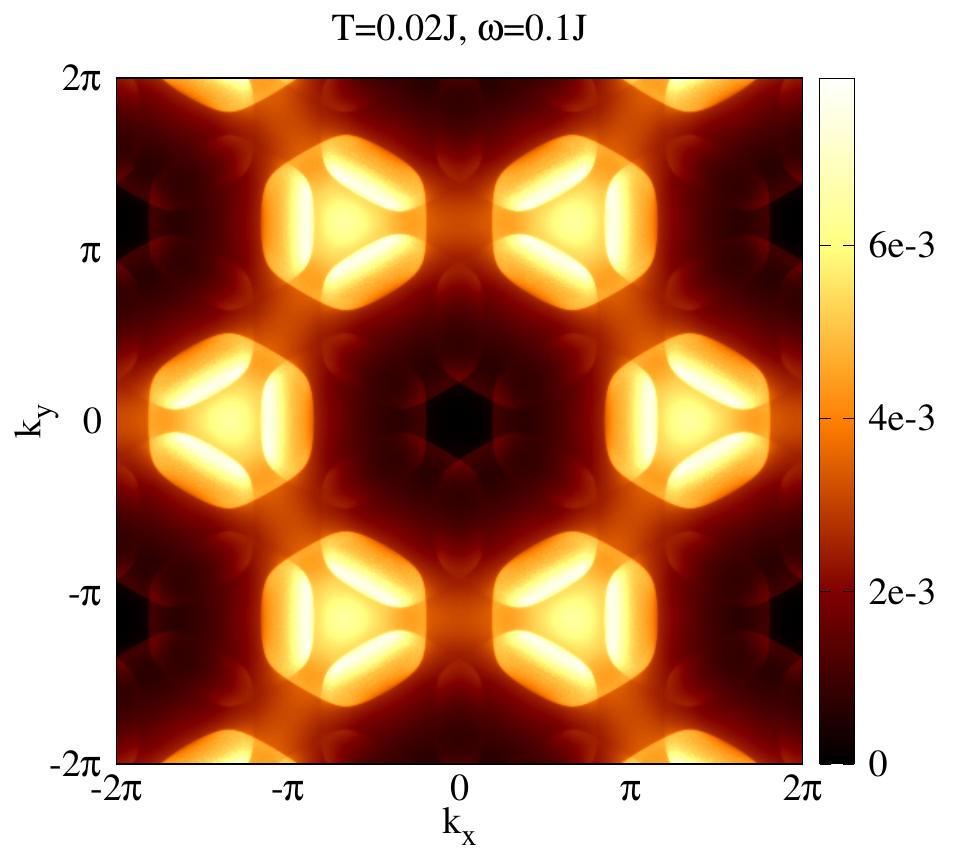}\\
\hspace{-0.15 cm}
\includegraphics[width=.348\textwidth]{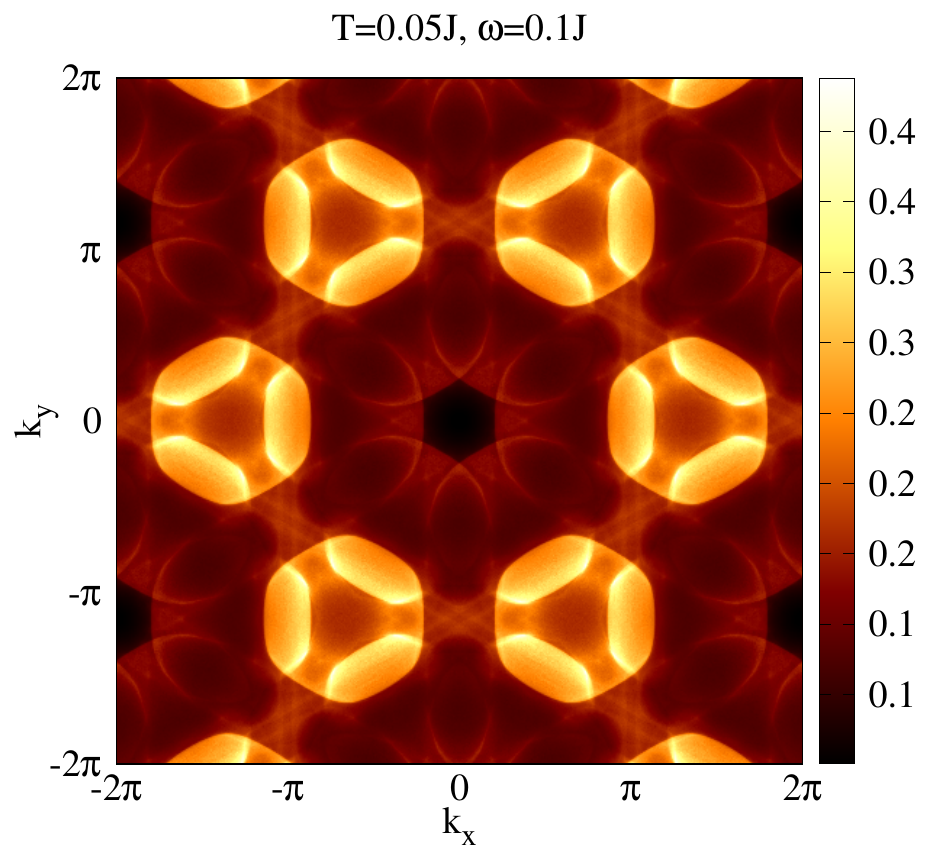}
\hspace{-0.15 cm}
\includegraphics[width=.348\textwidth]{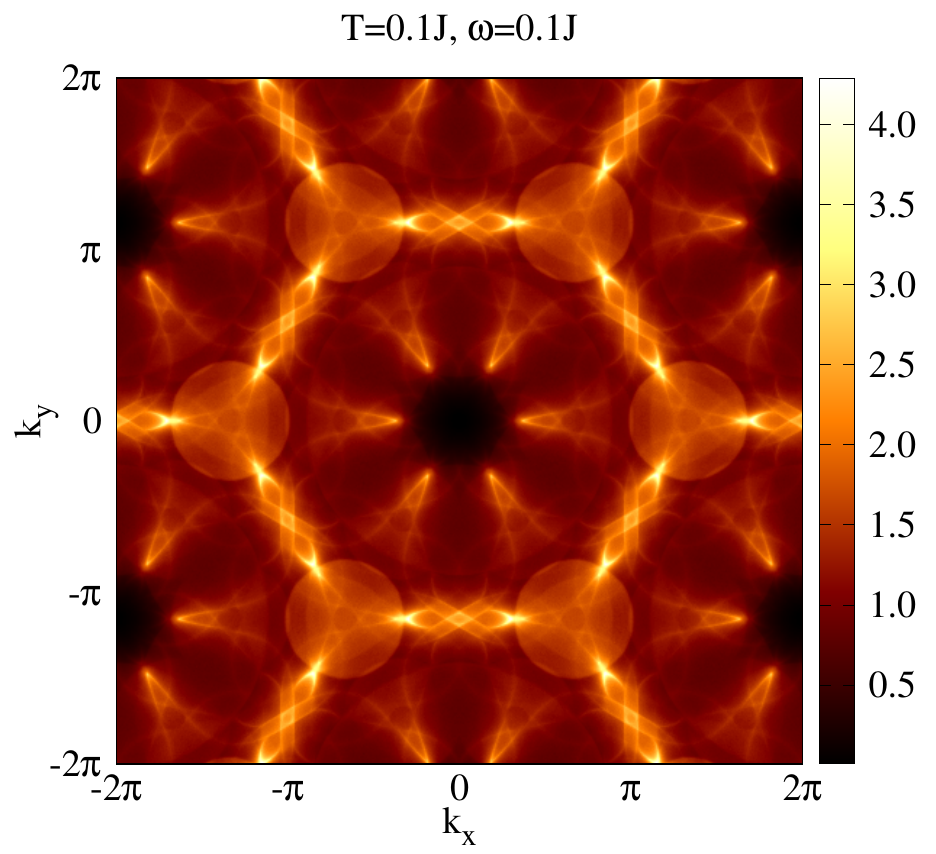}
\caption{(Color online). Same as Fig.~\ref{fig:DSF_q0_omegaLow} but for the $\sqrt{3}\times\sqrt{3}$ Ansatz.}
\label{fig:DSF_Sqrt3_omegaLow} 
\end{figure*}

The finite-temperature DSF can be understood by thinking of an INS experiment, where the incoming neutron exchanges with the system a net momentum $\mathbf{k}$ and a net energy $\omega$. As in the zero-temperature case, an incoming neutron can transfer a net momentum $\mathbf{k}$ and a net energy $\omega\geq0$ if and only if there are two spinons whose eigenenergies sum to $\omega$ at momenta that sum to $\mathbf{k}$. At finite temperature, on the other hand, the spinons are thermally excited, and thus they can also transfer net energy (in such a case $\omega<0$) to the neutron. Moreover, the net energy exchange at finite temperature can either be sums or differences, giving rise to the first three terms in~\eqref{eq:DSF}, in addition to the fourth that is the only remaining term at zero temperature. Indeed, in the limit $T\to0$,~\eqref{eq:DSF} reduces to the zero-temperature DSF derived in Ref.~\onlinecite{Halimeh2016}.

What is particularly interesting about~\eqref{eq:DSF} is that terms that only appear at finite temperature are not all exponentially suppressed by the spin gap $2\Delta$. In fact, two terms are exponentially suppressed only by the spinon gap $\Delta$, and thus it would be interesting to see if these terms will lead to substantial contributions at low temperature. Of course, this will actually also depend on the numerical values of the factors $\mathscr{B}^{r,s,m,n}_{\mathbf{k},\mathbf{q}}$ and $\mathscr{C}^{r,s,m,n}_{\mathbf{k},\mathbf{q}}$, and can provide clear signature of deconfinement of spinons.

\section{Results and discussion}\label{sec:results}

Numerically calculating~\eqref{eq:DSF} in the presence of Dirac-delta functions is problematic due to the zero support these functions have. As such, we approximate the Dirac-delta functions in~\eqref{eq:DSF} by Lorentzians with width $10^{-3}$, and subsequently use the VEGAS\cite{Lepage1978} Monte Carlo integration routine to numerically evaluate the finite-temperature DSF, which has proven to be a viable scheme in previous works.\cite{Punk2014,Halimeh2016} In all our numerical calculations, the spin length is set to $\mathcal{S}=0.2$, and we use the self-consistent parameters shown in Fig.~\ref{fig:fields}. The choice of $\mathcal{S}=0.2$ is to ensure that we are in the quantum spin liquid phase,\cite{Sachdev1992} and additionally serves to provide continuity with previous work.\cite{Halimeh2016}

\begin{figure}[t]
\centering
\hspace{-0.15 cm}
\includegraphics[width=.351\textwidth]{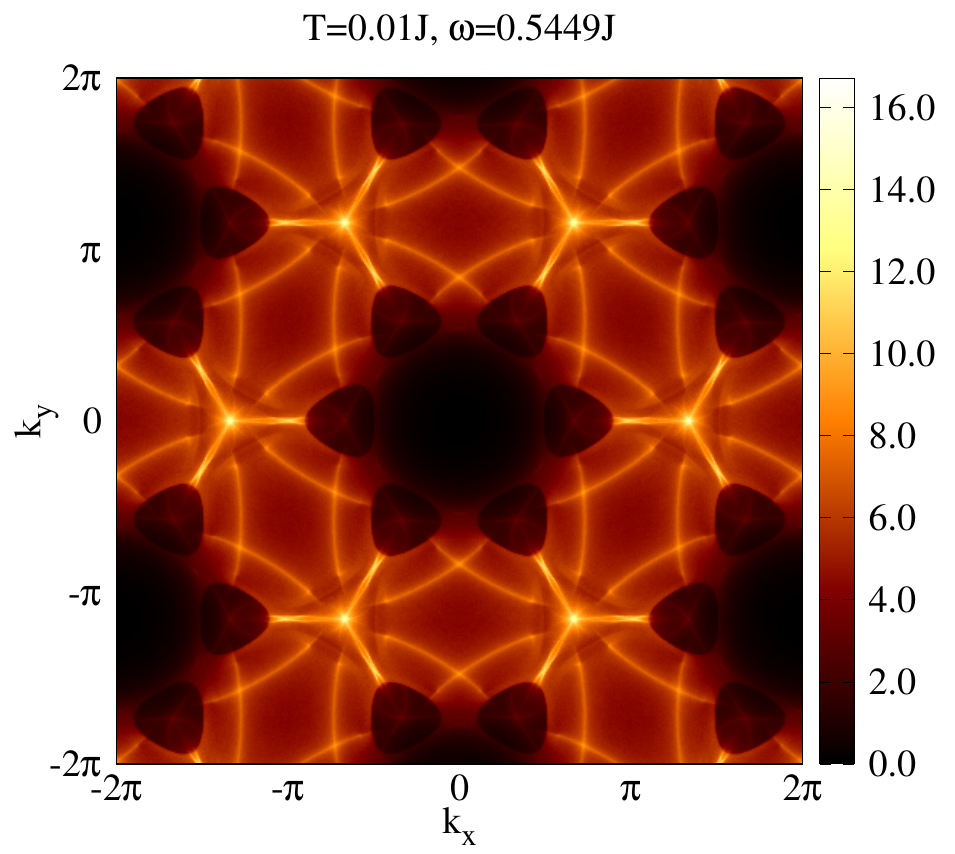}
\hspace{-0.15 cm}
\includegraphics[width=.351\textwidth]{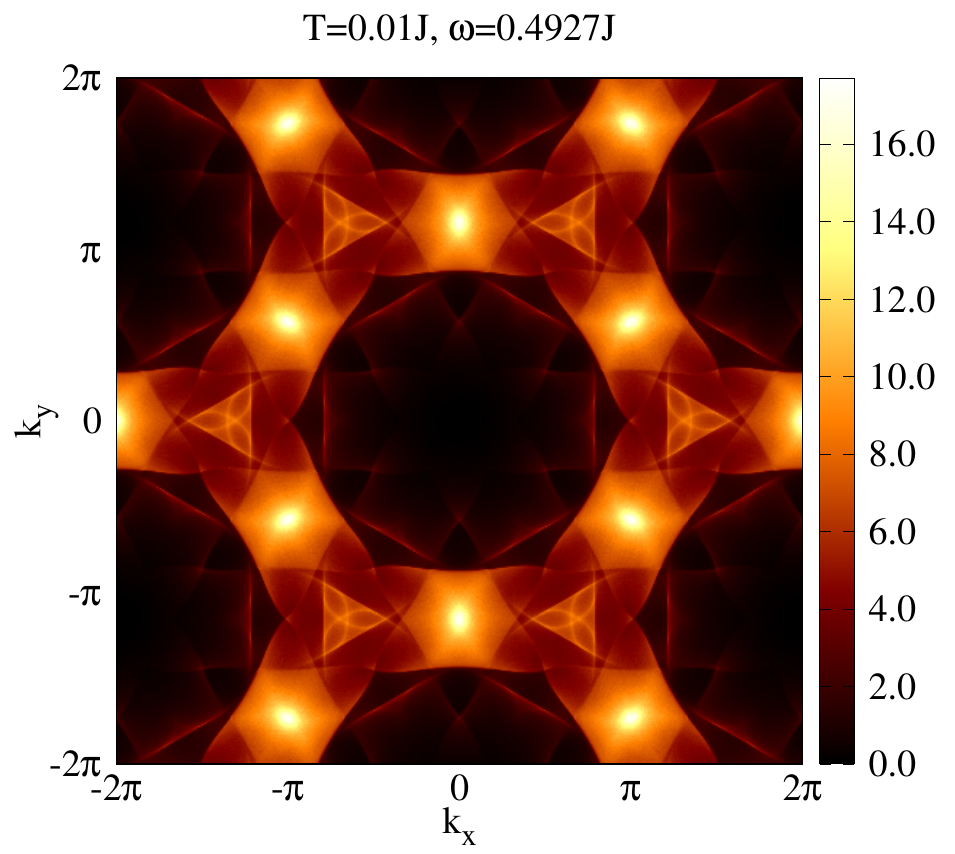}
\caption{(Color online). DSF results at high energies for the $\mathbf{q}=0$ (top panel) and $\sqrt{3}\times\sqrt{3}$ (bottom panel) Ans\"atze. Unlike at lower energies (cf.~Figs.~\ref{fig:DSF_q0_omegaLow} and~\ref{fig:DSF_Sqrt3_omegaLow}), the DSF changes insignificantly and is still almost the same up to $T=0.1J$.} 
\label{fig:DSF_omegaHigh}
\end{figure}

\begin{figure*}[htp]
 \centering
 \includegraphics[width=0.7\textwidth]{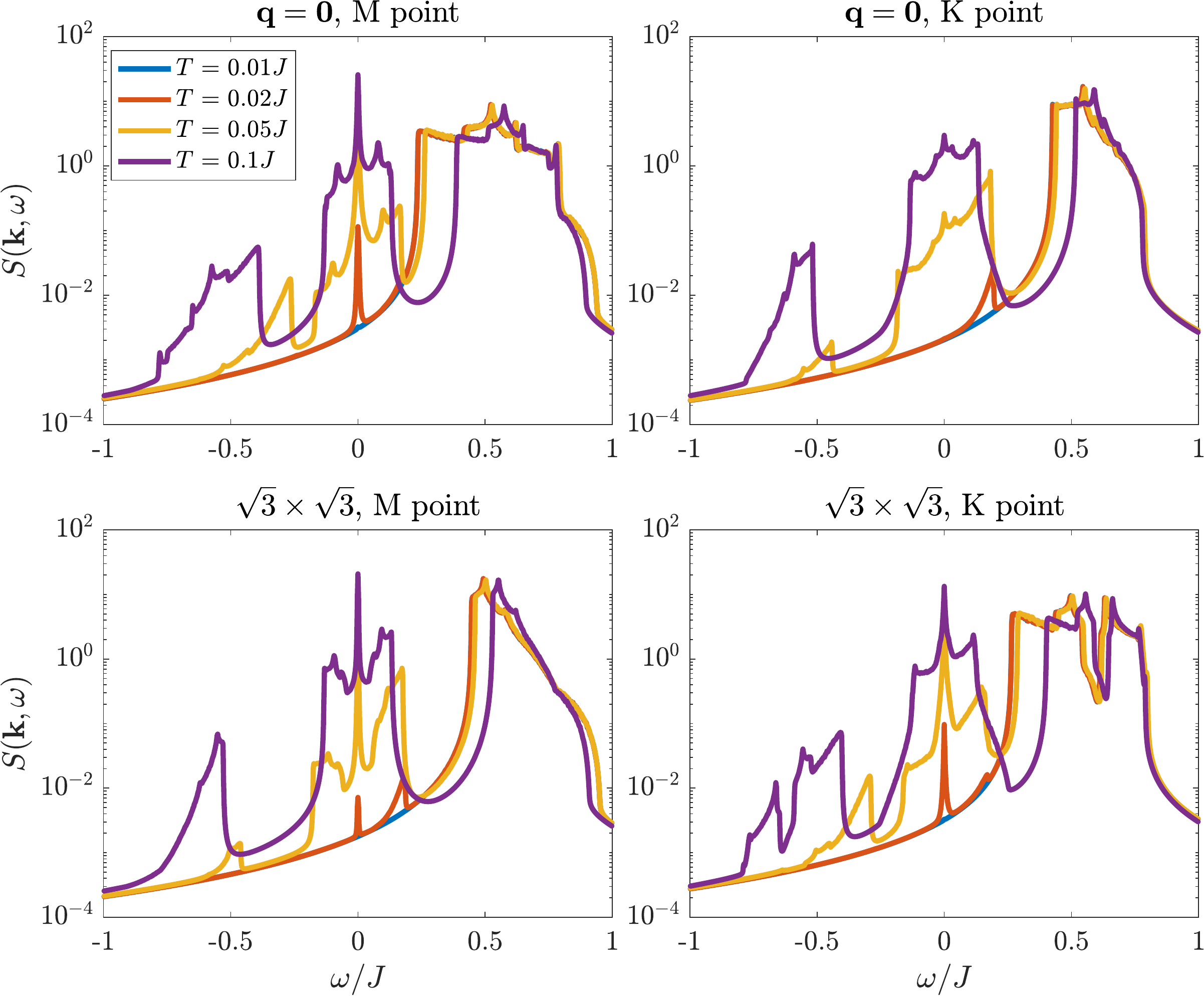}
 \caption{(Color online). The DSF for the Ans\"atze under consideration at the K and M points as function of frequency. The spin gap shows rapid filling with temperature.
 }
 \label{fig:wavevector}
\end{figure*}

We begin with the finite-temperature DSF results shown in Fig.~\ref{fig:DSF_q0} for the $\mathbf{q}=\bm{0}$ Ansatz along the $\Gamma$-M-K-$\Gamma$ high-symmetry lines at low temperatures $T\leq0.1J$. Even though at zero-temperature there is no spectral weight at all below the spin gap in the DSF, we see that even at very small temperature $T=0.01J$ there is already nonnegligible spectral weight filling up the spin gap continuously down to negative frequencies. This spectral weight arises from the first three terms in~\eqref{eq:DSF}, which completely vanish at $T=0J$. Physically in an INS setup, this means that due to thermal excitations, processes exist where the incoming neutron and an excited spinon impart (absorb) energy on (from) a second spinon, which gives rise to weight in the DSF at positive (negative) frequencies that are smaller than the spin gap in magnitude. Also, this can alternatively mean that two excited spinons impart energy on the incoming neutron, which contributes weight only at negative frequencies $\omega\leq-2\Delta$ in the DSF. Even though it seems that below the spin gap the DSF is homogeneous over momentum at $T=0.01J$, Fig.~\ref{fig:DSF_q0_omegaLow} shows that at this temperature at fixed frequency $\omega=0.1J$, the DSF has a rich structure with minimum at the $\Gamma$ point $(k_x,k_y)=(0,0)$ and maximum at the M point $(k_x,k_y)=(0,2\pi/\sqrt{3})$, and with the K point $(k_x,k_y)=(2\pi/3,2\pi/\sqrt{3})$ being of an intermediate spectral weight. This is remarkably similar to the INS measurement of Han \textit{et al.}~at $T\sim J/100$ and $\omega\sim J/10$ in Fig.~1(c) of Ref.~\onlinecite{Han2012}. However, unlike their result, we do not find that the DSF is constant as a function of frequency. In fact, around the spin gap, the DSF is about three orders of magnitude larger than at $\omega=0.1J$.

As the temperature is increased to $T=0.02J$, the spectral weight around $\omega=0.1J$ is already larger by almost a factor of three from what it is at $T=0.01J$ as can be seen in Figs.~\ref{fig:DSF_q0} and~\ref{fig:DSF_q0_omegaLow}. Interestingly, the DSF at $\omega=0.1J$ in Fig.~\ref{fig:DSF_q0_omegaLow} shows a notable change at $T=0.02J$ compared to $T=0.01J$, whereas the maximum at the M point in the latter now shows a hexagram structure of lower spectral intensity. A similar hexagram structure of yet lower intensity also appears at the $\Gamma$ point.

At $T=0.05J$, the spectral weight around $\omega=0.1J$ is over two orders of magnitude larger than at $T=0.01J$ at the same frequency, whereby the signal in the DSF around zero frequency shown in Fig.~\ref{fig:DSF_q0} compares in weight to that above the spin gap in certain regions. Interestingly, we see that at $\omega=0J$ the weight concentrates at the $\Gamma$ and M points even though the highest-intensity point over the whole DSF is at the K point at roughly $\omega=0.5532J$. Note that at zero temperature, there is no weight at all at the $\Gamma$ point, and this is due to the fact that the ground state has a total spin of zero. At finite temperature, there are thermal excitations and the system does not have zero total spin. The DSF at $\omega=0.1J$ for this temperature is also given in Fig.~\ref{fig:DSF_q0_omegaLow}, where its structure is similar to that at $T=0.02J$.

As the temperature is increased to $T=0.1J$, $\Gamma$ becomes the highest-intensity point in the DSF, and the M point at zero frequency overtakes in intensity the K point at $\omega=0.5532J$. Moreover, it can be seen that the DSF seems to be splitting into three distinct separate regions, one at positive frequency, a second at negative frequency, and a third region around $\omega=0J$. From~\eqref{eq:DSF} it is easy to determine which terms contribute to each region. The first term in~\eqref{eq:DSF} is responsible for the DSF weight at negative frequencies, and this becomes more prominent with higher temperature, as the spinons are more thermally excited and hence it is more likely that two spinons impart their energy on the incoming neutron. The second and third terms of~\eqref{eq:DSF} correspond to processes where a neutron and an excited spinon impart (absorb) energy on (from) a second spinon, leading to the extended region around zero frequency in the DSF. This contribution also grows with temperature. The DSF at $\omega=0.1J$ shown in Fig.~\ref{fig:DSF_q0_omegaLow} is also significantly different from that shown at lower temperatures. We remark that at this temperature, the spin density is still small enough such that interactions may be neglected and SBMFT therefore remains valid, but as the lower panel of Fig.~\ref{fig:SD} clarifies, here we are in a regime where the spinon density is increasing rapidly, and thus SBMFT cannot be fully trusted at any higher temperatures. Indeed, this three-region structure of the DSF becomes even more prominent at higher temperatures. A discussion thereof is provided in Appendix~\ref{sec:DSF_highT}.

Note that all the structures in Fig.~\ref{fig:DSF_q0_omegaLow} exhibit a sixfold rotation symmetry around the $\Gamma$ point due to the symmetric nonchiral nature of the $\mathbf{q}=\bm{0}$ Ansatz where time-reversal symmetry is preserved. In the case of chiral Ans\"atze such as cuboc1,\cite{Messio2012} the DSF displays time-reversal symmetry breaking through a reduction of the sixfold rotation symmetry around the $\Gamma$ point to a threefold one, whereas the static spin structure factor (SSF) is always invariant under $\mathbf{k}\to-\mathbf{k}$.\cite{Halimeh2016} For the latter, see Appendix~\ref{sec:SSF} for examples.

In addition to our results for the $\mathbf{q}=\bm{0}$ Ansatz, we also calculate in Fig.~\ref{fig:DSF_Sqrt3} the DSF for the $\sqrt{3}\times\sqrt{3}$ Ansatz along the $\Gamma$-M-K-$\Gamma$ high-symmetry lines. The same behavior manifests itself as in the case of the $\mathbf{q}=\bm{0}$ Ansatz. As temperature is increased, the spin gap of the DSF is rapidly filled with spectral weight even when the temperature is much lower that the spin gap itself. We also present the DSF for the $\sqrt{3}\times\sqrt{3}$ Ansatz at fixed frequency $\omega=0.1J$ in Fig.~\ref{fig:DSF_Sqrt3_omegaLow}, where we see that, just as in the case of the $\mathbf{q}=\bm{0}$ Ansatz, the structure of the DSF is very rich even at very low $T$, and it changes noticeably as the temperature is increased. Also as in the case of the $\mathbf{q}=\bm{0}$ Ansatz, at $T=0.1J$ a three-region structure emerges in the DSF seen in Fig.~\ref{fig:DSF_Sqrt3}. This facet is further discussed in Appendix~\ref{sec:DSF_highT}. At zero temperature, the DSF of the $\sqrt{3}\times\sqrt{3}$ Ansatz has its highest intensity at the M point, and yet with increasing temperature, we see that at $\omega=0J$ the K point has more weight than the M point. This is similar to the case of the $\mathbf{q}=\bm{0}$ Ansatz but with the points interchanged. We remark that even though it is relatively easy to tell both Ans\"atze apart from their DSF at the lower temperatures, the distinction is much less obvious at higher temperatures. Indeed, in Appendix~\ref{sec:DSF_highT} the DSF is basically identical for both at $T=0.19J$ when SBMFT implies a phase close to a trivial paramagnet, but the theory is unreliable at such high temperatures due to the significant spinon density; cf.~bottom panel of Fig.~\ref{fig:SD}.

In Fig.~\ref{fig:DSF_omegaHigh}, we show the DSF at $T=0.01J$ at high frequency for both Ans\"atze. Once again, the DSF exhibits six-fold rotation symmetry around the $\Gamma$ point due to time-reversal symmetry. We note that we also calculate this DSF at higher temperatures $T\leq0.1J$ but we do not present these results as they look almost identical to their $T=0.01J$ counterparts besides a faint smoothening effect. As a further probe of the frequency dependence in the DSF, we plot it for each Ansatz in Fig.~\ref{fig:wavevector} for the K and M points over the frequency range $\omega/J\in[-1,1]$. In accordance with our description above, we see that the spectral weight around zero frequency is much smaller than at the spin-gap energy for $T=0.01J$, although nonnegligible given our numerical accuracy and the rich structures in Figs.~\ref{fig:DSF_q0_omegaLow} and~\ref{fig:DSF_Sqrt3_omegaLow}. However, the spin gap quickly fills up with temperature, with a significant zero-frequency peak already at $T=0.02J$ for the M point in both Ans\"atze. By $T=0.05J$, the spectral weight around the zero-frequency region is almost of the same order as that at the spin-gap energy. We again see the three-region structure forming in the DSF at $T=0.1J$, which we have already discussed. Also as previously mentioned, we see that even though for the $\mathbf{q}=\bm{0}$ ($\sqrt{3}\times\sqrt{3}$) Ansatz the K (M) point is always the highest in spectral weight over the entire DSF at very low temperature, as the temperature is raised, the zero-frequency spectral weight builds more intensely at the M (K) point. It is also worth mentioning that the frequency-dependent nature of the DSF as shown in Fig.~\ref{fig:wavevector} is in contrast to INS measurements\cite{Han2012} that show the DSF to be constant as a function of $\omega$ -- apart from the peak at the $\Gamma$ point, which most likely is due to dirt in the sample. Nevertheless, such $\omega$-dependence can be vastly removed by including spinon-vison interactions that lead to a structureless DSF.\cite{Punk2014} In fact, in Ref.~\onlinecite{Punk2014} spinon-vison interactions do not succeed in completely removing an onset in the DSF, where one still remains at low frequency. Our results show that this onset is completely removed even at quite low temperatures. Therefore, we expect that a finite-temperature extension of Ref.~\onlinecite{Punk2014} would bring the numerical and experimental results for the DSF to great agreement. This is beyond the scope of the current paper, however, and we leave it open for future work. We summarize the finite-temperature contribution to the DSF in Table~\ref{Table}. Even though the contribution is very small for $T=0.01J$, we find that it is more than four percentage points at $T=0.05J$ where SBMFT is expected to still be reliable.

From a different point of view, our results rely on an SBMFT self-consistently determined spinon gap that is known to be an overestimate of its actual physical value. In fact, in Ref.~\onlinecite{Punk2014} this is taken into account by setting the gap to a value smaller than its self-consistent result. In our case, this is something that we can also do in principle. For example, if we are at temperature $T$ and decrease the gap by a factor of two, we would see the same level of spin-gap filling happening originally at $2T$; cf.~\eqref{eq:DSF}. This in principle would bring our results qualitatively even closer to the measurements of Ref.~\onlinecite{Han2012}. Similarly, our results nontrivially depend on the value of $\mathcal{S}$, which we have set to $0.2$ due to continuity with previous work and to ensure that we are deep in the quantum regime. Indeed, if we increase $\mathcal{S}$, this would actually decrease our spinon gap,\cite{Sachdev1992,Kos2017} eventually closing the gap and forming a condensate as long-range order emerges. Therefore, our results would be even further enhanced at larger $\mathcal{S}$ where a smaller finite spinon gap arises.

Importantly, we note that we have checked that our results obey the sum rule\cite{Auerbach1994} (see Appendix~\ref{sec:SSF}), and additionally verified that the finite-temperature DSF satisfies the relation of detailed balance (for an example, see Appendix~\ref{sec:DB}).

\begin{table}[]
	\centering
	\caption{Self-consistently calculated spin gap $2\Delta$ for the $\mathbf{q}=\bm{0}$ SBMFT Ansatz on the AFKM as a function of temperature, along with the contribution percentage $\mathfrak{f}$ to the DSF from energies $\omega<2\Delta$.}
	\medskip
	\begin{tabular}{|c"c|c|c|c|c}%{@{}|l|lllll@{}}
		\bottomrule
		\thickhline
		
		$T/J$ &  \multicolumn{1}{c|}{$0$} &  \multicolumn{1}{c|}{$0.01$} &  \multicolumn{1}{c|}{$0.02$} &  \multicolumn{1}{c|}{$0.05$} &  \multicolumn{1}{c|}{$0.1$} \\ \cmidrule(r){1-1}
		\hline
		$\mathcal{A}$ & \multicolumn{1}{c|}{$0.26269$} & \multicolumn{1}{c|}{$0.26269$} & \multicolumn{1}{c|}{$0.26268$} & \multicolumn{1}{c|}{$0.26058$}  & \multicolumn{1}{c|}{$0.23357$}  \\ \cmidrule(r){1-1}
		\hline
		$-\mathcal{B}$ & \multicolumn{1}{c|}{$0.05729$} & \multicolumn{1}{c|}{$0.05729$} & \multicolumn{1}{c|}{$0.05730$} & \multicolumn{1}{c|}{$0.05682$}  & \multicolumn{1}{c|}{$0.04373$}  \\ \cmidrule(r){1-1}
		\hline
		$\lambda$ & \multicolumn{1}{c|}{$0.41268$} & \multicolumn{1}{c|}{$0.41268$} & \multicolumn{1}{c|}{$0.41270$} & \multicolumn{1}{c|}{$0.41319$}  & \multicolumn{1}{c|}{$0.40498$}  \\ \cmidrule(r){1-1}
		\hline
		$2\Delta/J$ & \multicolumn{1}{c|}{$0.26296$} & \multicolumn{1}{c|}{$0.26458$} & \multicolumn{1}{c|}{$0.26480$} & \multicolumn{1}{c|}{$0.28851$}  & \multicolumn{1}{c|}{$0.40188$}  \\ \cmidrule(r){1-1}
		\hline
		$\mathfrak{f}(\%)$ & \multicolumn{1}{c|}{$0$} & \multicolumn{1}{c|}{$5.79\times10^{-5}$} & \multicolumn{1}{c|}{$4.34\times10^{-2}$} & \multicolumn{1}{c|}{$4.32$}  & \multicolumn{1}{c|}{$30.57$}  \\ \cmidrule(r){1-1}
		\hline

		\hline
		\thickhline
	\end{tabular}
	\label{Table}
\end{table}

\section{Conclusion and outlook}\label{sec:conclusion}
In conclusion, we have analytically derived and numerically calculated in the framework of Schwinger-boson mean-field theory the static and dynamic spin structure factor at low temperatures of the spin-$1/2$ antiferromagnetic Heisenberg kagome model for two prominent symmetric Ans\"atze, the $\mathbf{q}=\bm{0}$ and the $\sqrt{3}\times\sqrt{3}$. Our numerical results show that the structure factors change qualitatively with increasing temperature, where the spin gap rapidly fills up with temperature. Moreover, this population of the spin gap in the DSF occurs already at temperatures more than an order of magnitude smaller than the spin gap itself, and before any significant changes in the mean-field parameters have occured, or the spinon density has nontrivially increased. This happens because there are terms in the low-frequency structure factor that are suppressed at finite temperatures by only $\exp(-\Delta/T)$ and is thus a clear signature of deconfinement of spinons. This may explain in part the results of INS experiments\cite{Han2012} where there is no onset of the two-spinon continuum even at temperatures of the order of $J/100$. A question that immediately presents itself in the wake of our results is whether finite temperature can bring full agreement between the theoretical results of Punk \textit{et al.}~in Ref.~\onlinecite{Punk2014} and the experimental measurements of Han \textit{et al.}~in Ref.~\onlinecite{Han2012}. Indeed, our results show that finite temperature completely removes any sharp onset in the DSF down to negative frequencies, but the DSF is still clearly frequency-dependent. The inclusion of spinon-vison interactions in Ref.~\onlinecite{Punk2014} leads to a DSF that is more or less structureless and flattened at low energies, but that still exhibits an onset at low frequencies. Hence, an extension of this study to finite temperature may significantly advance the agreement between theory and experiment. We have also discussed that since the SBMFT self-consistent spinon gap is actually larger than its physical value, we can use a smaller value in our numerical simulations such that the spin gap fills up more rapidly in the DSF at a given temperature, thereby bringing our results closer to what is observed experimentally in Ref.~\onlinecite{Han2012} even without including spinon-vison interactions.

We have additionally discussed the shortcomings of SBMFT at high temperatures, and explained how this leads to a three-region structure in the DSF due to the system spectrum approaching a quasi-elastic profile. The spinon density can be used as a guide as to when SBMFT is reliable, because so long as the density of spinons is very small, then interactions can be effectively neglected rendering SBMFT a good description of the system. As temperature is raised, the spin density rapidly rises, and then SBMFT results are no longer accurate. It would be especially interesting to extend SBMFT in order to be able to account for high temperatures where the nearest-neighbor correlations disappear. This would give us a platform to compare SBMFT results to those obtained in NLCE\cite{Sherman2018} for temperatures $T\geq J/4$, which is above what SBMFT can reliably describe. Another interesting study would be the behavior of chiral Ans\"atze at finite temperature, where it is expected that there would be a finite-temperature phase transition from a time-reversal symmetry broken phase at low temperature to a time-reversal symmetric phase at high temperature. This is the subject of an ongoing study by the current authors.

We here emphasize that the main conclusion of our work -- namely that at finite temperature the subgap spectral weight is suppressed only by $\exp(-\Delta/T)$ due to spinon deconfinement -- would still hold for other Ans\"atze than the ones discussed in this work. Indeed, as previously mentioned, a more accurate description of herbertsmithites would involve DM interactions. In SBMFT, this still involves a gapped $\mathbb{Z}_2$ spin liquid phase,\cite{Messio2017} and thus our qualitative result will still hold.

Finally, it is worth mentioning that this work, in using SBMFT, inherently assumes that the AFKM spin liquid phase is gapped. However, our results indicate that at finite temperature the debate over whether this phase is gapped or gapless may become irrelevant. Our results show that even at low temperatures there is nontrivial contribution to the DSF. Therefore, inelastic neutron scattering experiments would need to be at very small temperatures and very small energies -- both very challenging limits\cite{Han2016} -- in order to truly ascertain whether the AFKM QSL is gapped or gapless. We also emphasize that we are not saying that our results would be the entire explanation for the observed spectral weight at low energies (impurities could play a role, etc…) but our results have to be taken into account if this is really the physics of a gapped spin liquid.

\section*{Acknowledgments}
The authors are grateful to Felix Mackenroth for his help in Gnuplot; to Bernhard Frank, Johannes Lang, Jeffrey G.~Rau, and Simon Trebst for fruitful discussions; and to Paul A.~McClarty and Matthias Punk for valuable discussions and comments on our manuscript. The work of RRPS is supported in part by US National Science Foundation DMR grant number 1855111.

\appendix
\begin{figure*}[htp]
\centering
\hspace{-0.15 cm}
\includegraphics[width=.495\textwidth]{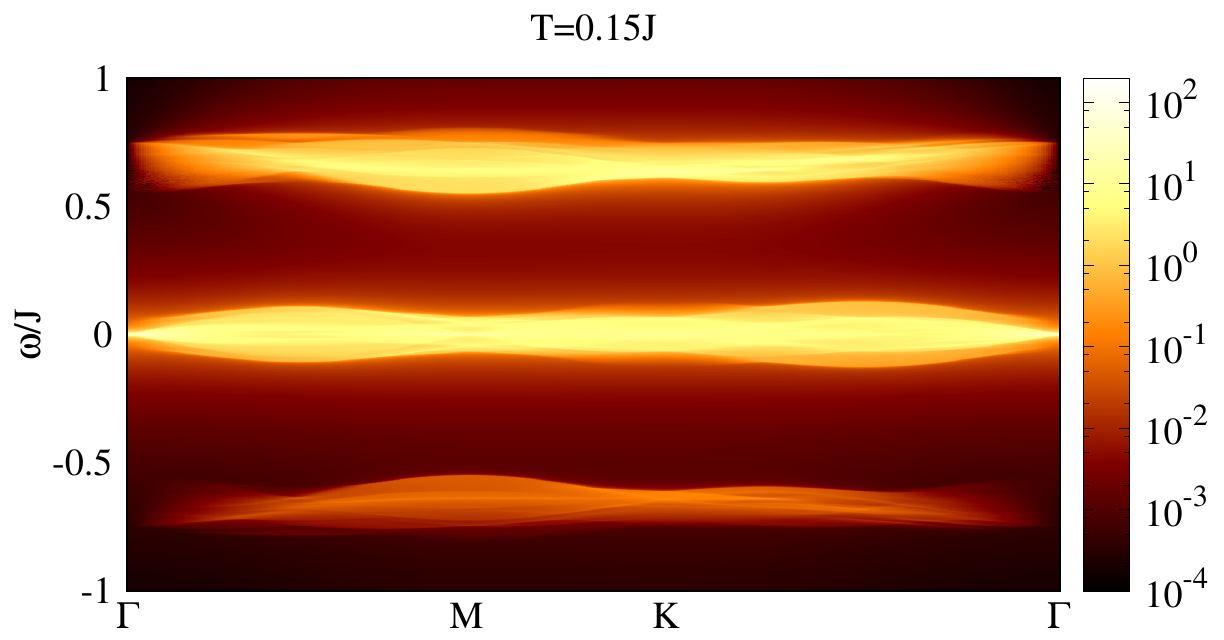}
\hspace{-0.15 cm}
\includegraphics[width=.495\textwidth]{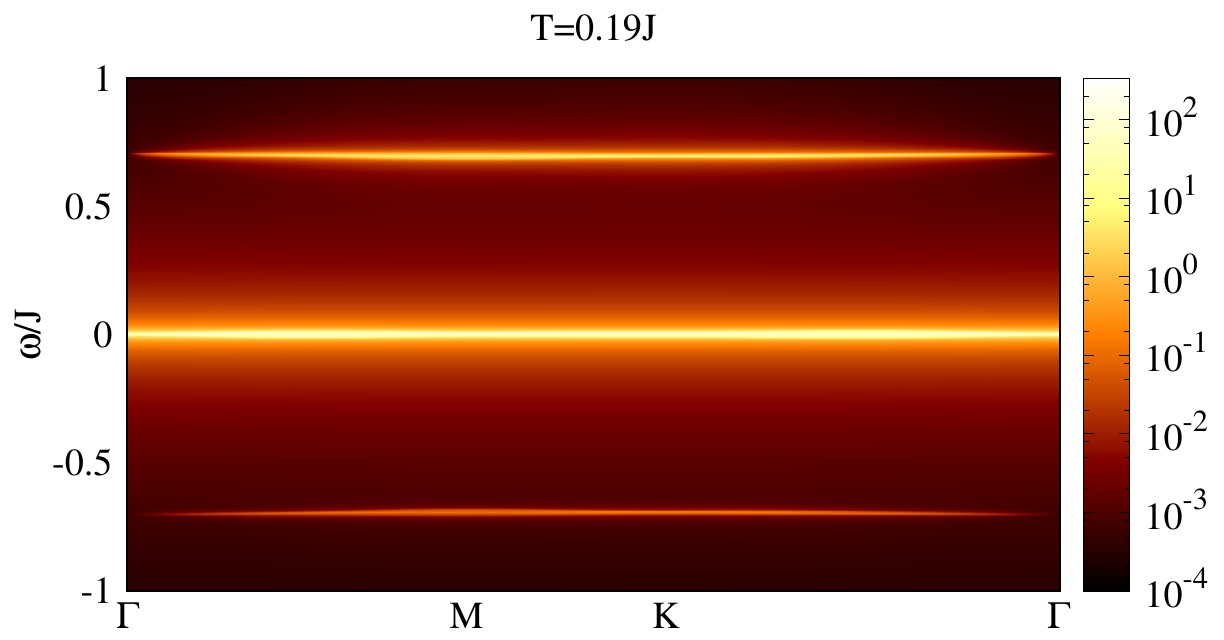}\\
\hspace{-0.15 cm}
\includegraphics[width=.495\textwidth]{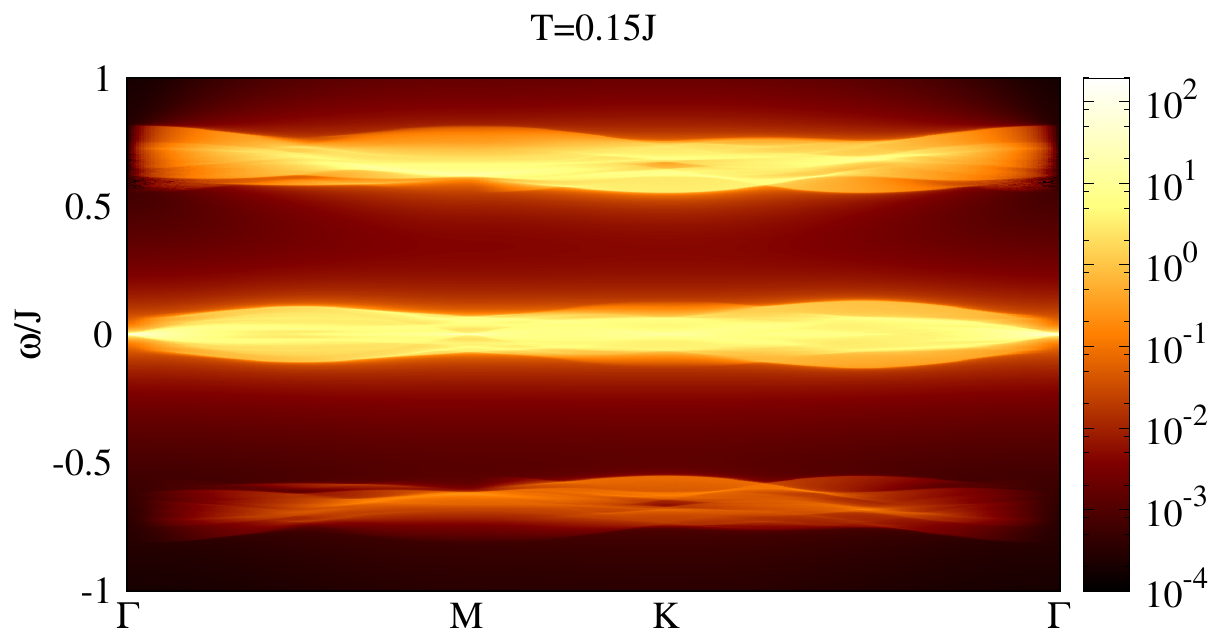}
\hspace{-0.15 cm}
\includegraphics[width=.495\textwidth]{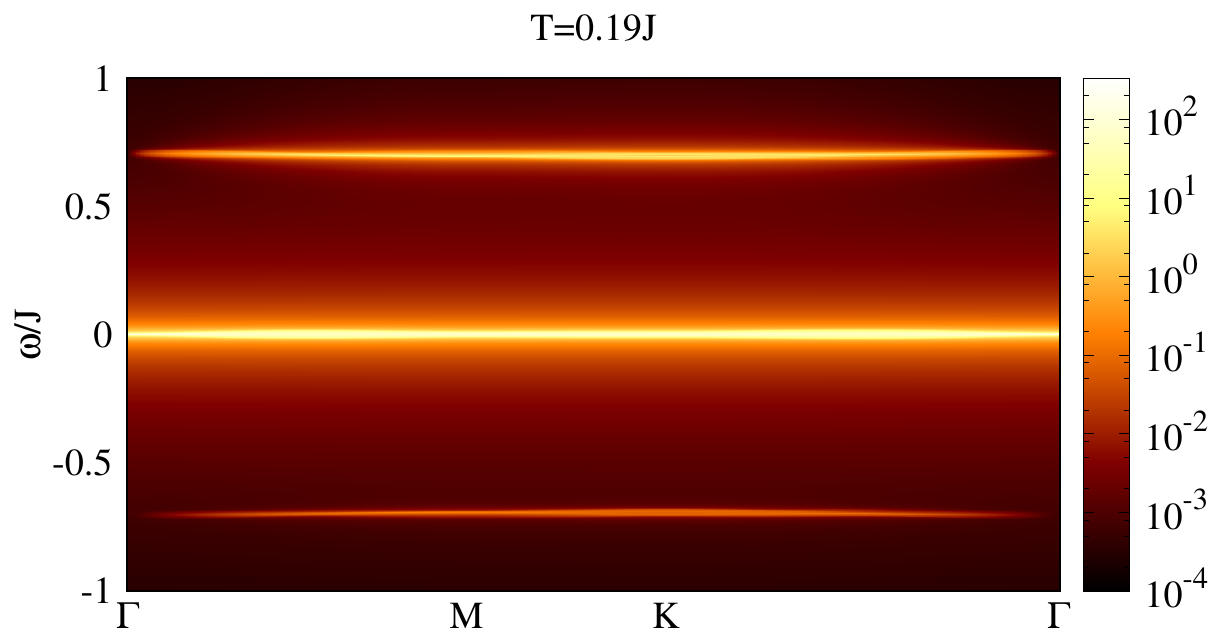}
\caption{(Color online). The dynamic spin structure factor for the $\mathbf{q}=\bm{0}$ (top panels) and the $\sqrt{3}\times\sqrt{3}$ (bottom panels) Ans\"atze along the $\Gamma$-M-K-$\Gamma$ high-symmetry lines at temperatures $T/J=0.15$ and $0.19$. The rich structure at lower temperatures is reduced to three high-intensity lines around $\omega=0$ and $\pm2\lambda$ at these high temperatures. This is due to the system approaching a scenario where all spinon bands are degenerate with eigenvalue $\lambda$. SBMFT results are not fully reliable here, as the spinon density is not small (cf.~Fig.~\ref{fig:SD}).}
\label{fig:DSF_highT} 
\end{figure*}

\begin{figure*}[htp]
\centering
\hspace{-0.15 cm}
\includegraphics[width=.3\textwidth]{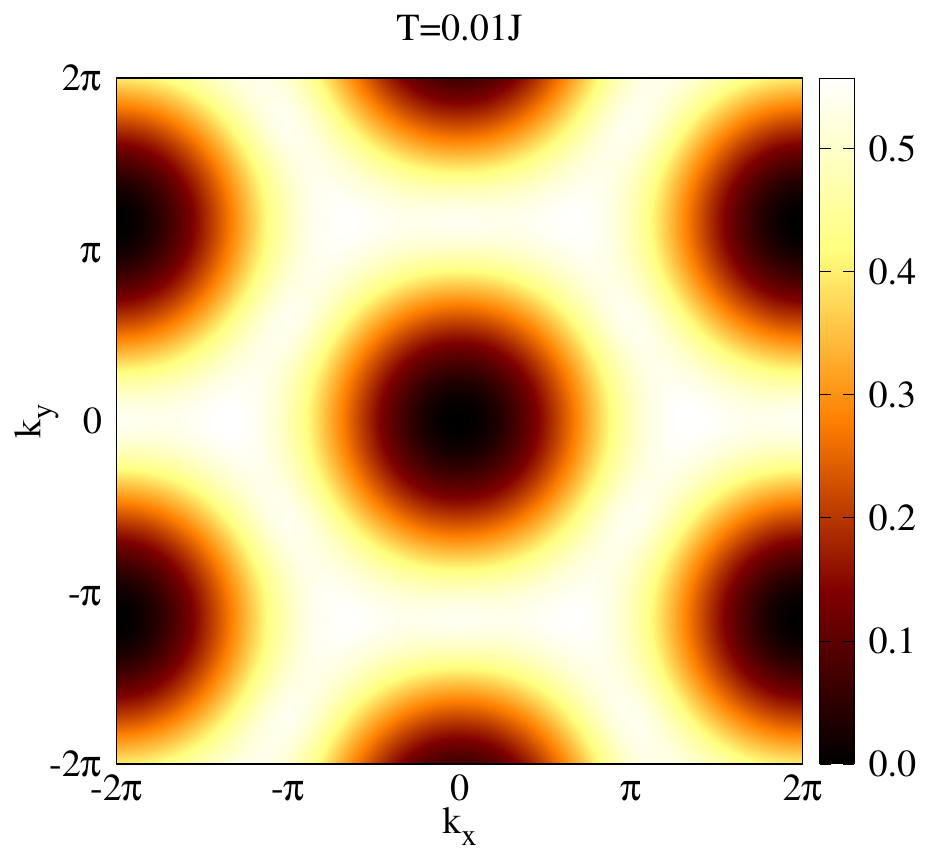}
\hspace{-0.15 cm}
\includegraphics[width=.3\textwidth]{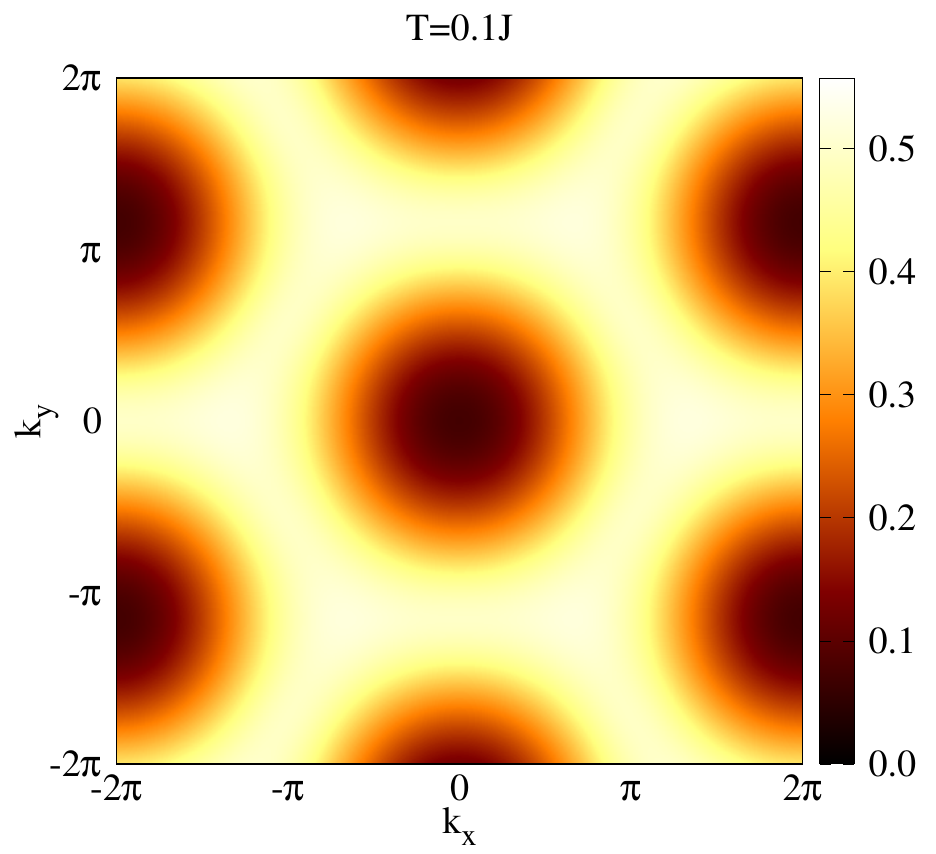}\\
\hspace{-0.15 cm}
\includegraphics[width=.3\textwidth]{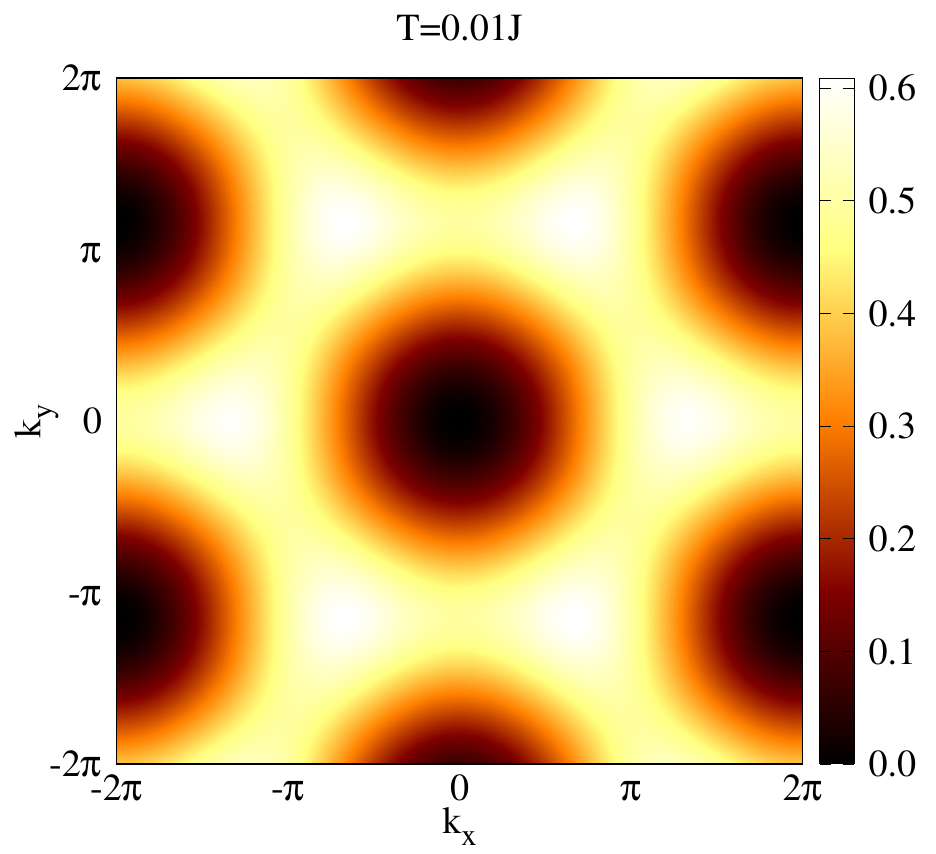}
\hspace{-0.15 cm}
\includegraphics[width=.3\textwidth]{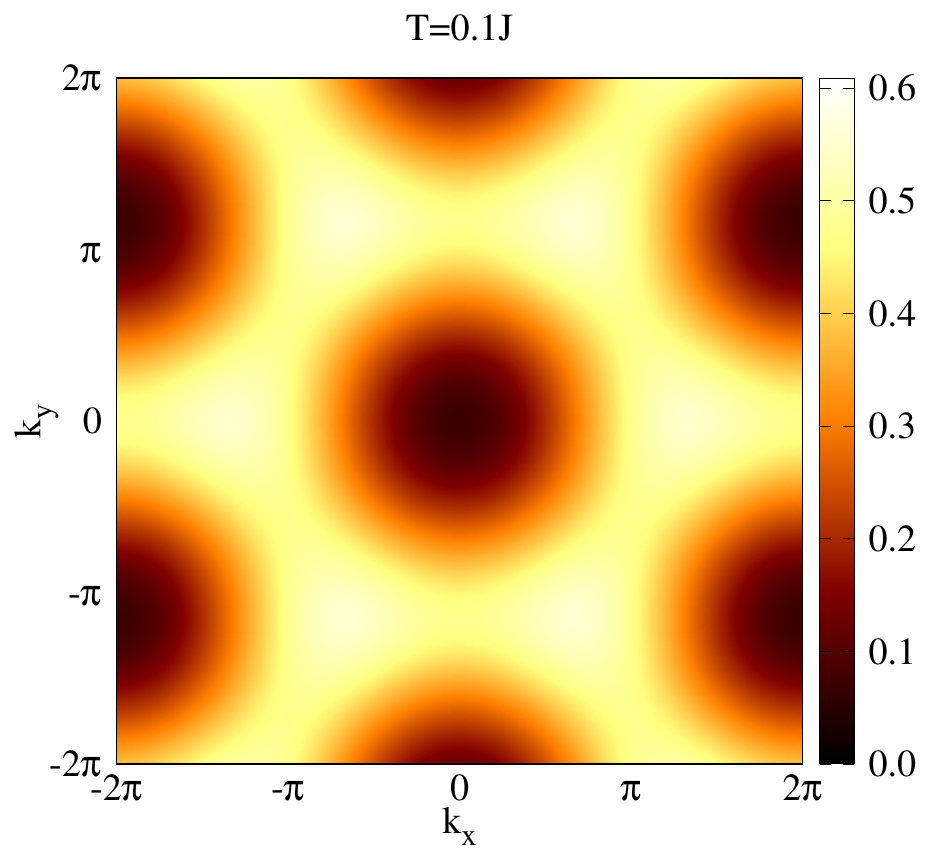}
\caption{(Color online). SSF for the $\mathbf{q}=0$ (top panels) and $\sqrt{3}\times\sqrt{3}$ (bottom panels) Ans\"atze, where we see that with temperature the structure of the SSF appears unaltered except for a small smoothening effect.}
\label{fig:SSF}
\end{figure*}
\section{Exact expressions for DSF terms}\label{sec:DSFterms}
Due to their length, the full expressions of the terms in~\eqref{eq:DSF} are provided here. The term responsible for processes where two thermally excited spinons impart energy on an incoming neutron is given by
\begin{widetext}
\begin{align}\nonumber
\mathscr{A}^{r,s,m,n}_{\mathbf{k},\mathbf{q}}=&\,U_{rm}^*(\mathbf{q})X_{rn}(\mathbf{k}+\mathbf{q})\left[X_{sn}^*(\mathbf{k}+\mathbf{q})U_{sm}(\mathbf{q})-V_{sm}(\mathbf{q})Y_{sn}^*(\mathbf{k}+\mathbf{q})\right]\\\nonumber
&+2U_{rm}^*(\mathbf{q})V_{rn}^*(-\mathbf{k}-\mathbf{q})\left[V_{sm}(\mathbf{q})U_{sn}(-\mathbf{k}-\mathbf{q})+V_{sn}(-\mathbf{k}-\mathbf{q})U_{sm}(\mathbf{q})\right]\\\nonumber
&+2Y_{rm}(-\mathbf{q})X_{rn}(\mathbf{k}+\mathbf{q})\left[X_{sm}^*(-\mathbf{q})Y_{sn}^*(\mathbf{k}+\mathbf{q})+X_{sn}^*(\mathbf{k}+\mathbf{q})Y_{sm}^*(-\mathbf{q})\right]\\\label{eq:term1}
&+Y_{rm}(-\mathbf{q})V_{rn}^*(-\mathbf{k}-\mathbf{q})\left[V_{sn}(-\mathbf{k}-\mathbf{q})Y_{sm}^*(-\mathbf{q})-X_{sm}^*(-\mathbf{q})U_{sn}(-\mathbf{k}-\mathbf{q})\right].
\end{align}
\end{widetext}
This is the term that is most suppressed in~\eqref{eq:DSF} with inverse temperature. The term that accounts for an incoming neutron imparting energy on two spinons, which is the only term that occurs at zero temperature, reads
\begin{widetext}
\begin{align}\nonumber
\mathscr{D}^{r,s,m,n}_{\mathbf{k},\mathbf{q}}=&\,X_{rm}^*(\mathbf{q})U_{rn}(\mathbf{k}+\mathbf{q})\left[U_{sn}^*(\mathbf{k}+\mathbf{q})X_{sm}(\mathbf{q})-Y_{sm}(\mathbf{q})V_{sn}^*(\mathbf{k}+\mathbf{q})\right]\\\nonumber
&+2X_{rm}^*(\mathbf{q})Y_{rn}^*(-\mathbf{k}-\mathbf{q})\left[Y_{sm}(\mathbf{q})X_{sn}(-\mathbf{k}-\mathbf{q})+Y_{sn}(-\mathbf{k}-\mathbf{q})X_{sm}(\mathbf{q})\right]\\\nonumber
&+2V_{rm}(-\mathbf{q})U_{rn}(\mathbf{k}+\mathbf{q})\left[U_{sm}^*(-\mathbf{q})V_{sn}^*(\mathbf{k}+\mathbf{q})+U_{sn}^*(\mathbf{k}+\mathbf{q})V_{sm}^*(-\mathbf{q})\right]\\\label{eq:term4}
&+V_{rm}(-\mathbf{q})Y_{rn}^*(-\mathbf{k}-\mathbf{q})\left[Y_{sn}(-\mathbf{k}-\mathbf{q})V_{sm}^*(-\mathbf{q})-U_{sm}^*(-\mathbf{q})X_{sn}(-\mathbf{k}-\mathbf{q})\right].
\end{align}
\end{widetext}
The terms of the DSF responsible for processes where a thermally excited spinon and the incoming neutron impart energy on a second spinon, or a thermally excited spinon imparts energy on a second spinon and the incoming neutron are
\begin{widetext}
\begin{align}\nonumber
\mathscr{B}^{r,s,m,n}_{\mathbf{k},\mathbf{q}}=&\,U_{rm}^*(\mathbf{q})U_{rn}(\mathbf{k}+\mathbf{q})\left[U_{sn}^*(\mathbf{k}+\mathbf{q})U_{sm}(\mathbf{q})-V_{sm}(\mathbf{q})V_{sn}^*(\mathbf{k}+\mathbf{q})\right]\\\nonumber
&+2U_{rm}^*(\mathbf{q})Y_{rn}^*(-\mathbf{k}-\mathbf{q})\left[V_{sm}(\mathbf{q})X_{sn}(-\mathbf{k}-\mathbf{q})+Y_{sn}(-\mathbf{k}-\mathbf{q})U_{sm}(\mathbf{q})\right]\\\nonumber
&+2Y_{rm}(-\mathbf{q})U_{rn}(\mathbf{k}+\mathbf{q})\left[Y_{sm}^*(-\mathbf{q})U_{sn}^*(\mathbf{k}+\mathbf{q})+V_{sn}^*(\mathbf{k}+\mathbf{q})X_{sm}^*(-\mathbf{q})\right]\\\label{eq:term2}
&+Y_{rm}(-\mathbf{q})Y_{rn}^*(-\mathbf{k}-\mathbf{q})\left[Y_{sn}(-\mathbf{k}-\mathbf{q})Y_{sm}^*(-\mathbf{q})-X_{sm}^*(-\mathbf{q})X_{sn}(-\mathbf{k}-\mathbf{q})\right],\\[1em]\nonumber
\mathscr{C}^{r,s,m,n}_{\mathbf{k},\mathbf{q}}=&\,X_{rm}^*(\mathbf{q})X_{rn}(\mathbf{k}+\mathbf{q})\left[X_{sn}^*(\mathbf{k}+\mathbf{q})X_{sm}(\mathbf{q})-Y_{sm}(\mathbf{q})Y_{sn}^*(\mathbf{k}+\mathbf{q})\right]\\\nonumber
&+2X_{rm}^*(\mathbf{q})V_{rn}^*(-\mathbf{k}-\mathbf{q})\left[X_{sm}(\mathbf{q})V_{sn}(-\mathbf{k}-\mathbf{q})+U_{sn}(-\mathbf{k}-\mathbf{q})Y_{sm}(\mathbf{q})\right]\\\nonumber
&+2V_{rm}(-\mathbf{q})X_{rn}(\mathbf{k}+\mathbf{q})\left[U_{sm}^*(-\mathbf{q})Y_{sn}^*(\mathbf{k}+\mathbf{q})+X_{sn}^*(\mathbf{k}+\mathbf{q})V_{sm}^*(-\mathbf{q})\right]\\\label{eq:term3}
&+V_{rm}(-\mathbf{q})V_{rn}^*(-\mathbf{k}-\mathbf{q})\left[V_{sn}(-\mathbf{k}-\mathbf{q})V_{sm}^*(-\mathbf{q})-U_{sm}^*(-\mathbf{q})U_{sn}(-\mathbf{k}-\mathbf{q})\right].
\end{align}
\end{widetext}
The two terms~\eqref{eq:term2} and~\eqref{eq:term3} are the ones responsible for the rapid filling of the spin gap in the DSF with temperature, while terms~\eqref{eq:term1} and~\eqref{eq:term4} contribute to the spectral weight in the DSF at $\omega\leq-2\Delta$ and at $\omega\geq2\Delta$, respectively.

As mentioned in the main text, at high temperatures the term~\eqref{eq:term2} dominates since the Bogoliubov matrices $V$ and $X$ are negligile. This gives rise to the dominance of the region around $\omega=0J$ at higher temperatures as seen in Fig.~\ref{fig:DSF_highT} in Appendix~\ref{sec:DSF_highT} below. Nevertheless, $V$ and $X$ are still finite, and this leads to two thin dimmer regions at around roughly $\omega=\pm2\lambda$ in Fig.~\ref{fig:DSF_highT}. It is to be noted that at such high temperatures where the spinon density is no longer small (cf.~Fig.~\ref{fig:SD}), SBMFT results cannot be fully trusted.

\section{DSF at highter temperatures}\label{sec:DSF_highT}

\begin{figure}[t]
 \centering
\hspace{-0.15 cm}
\includegraphics[width=.351\textwidth]{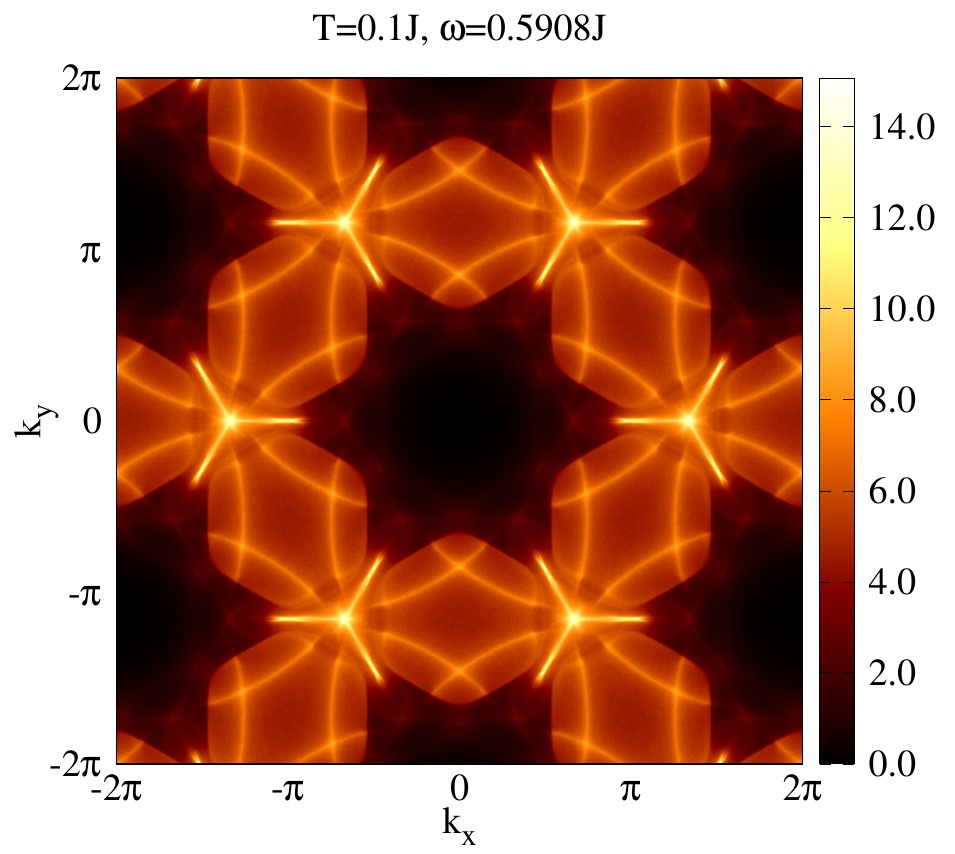}
\hspace{-0.15 cm}
\includegraphics[width=.351\textwidth]{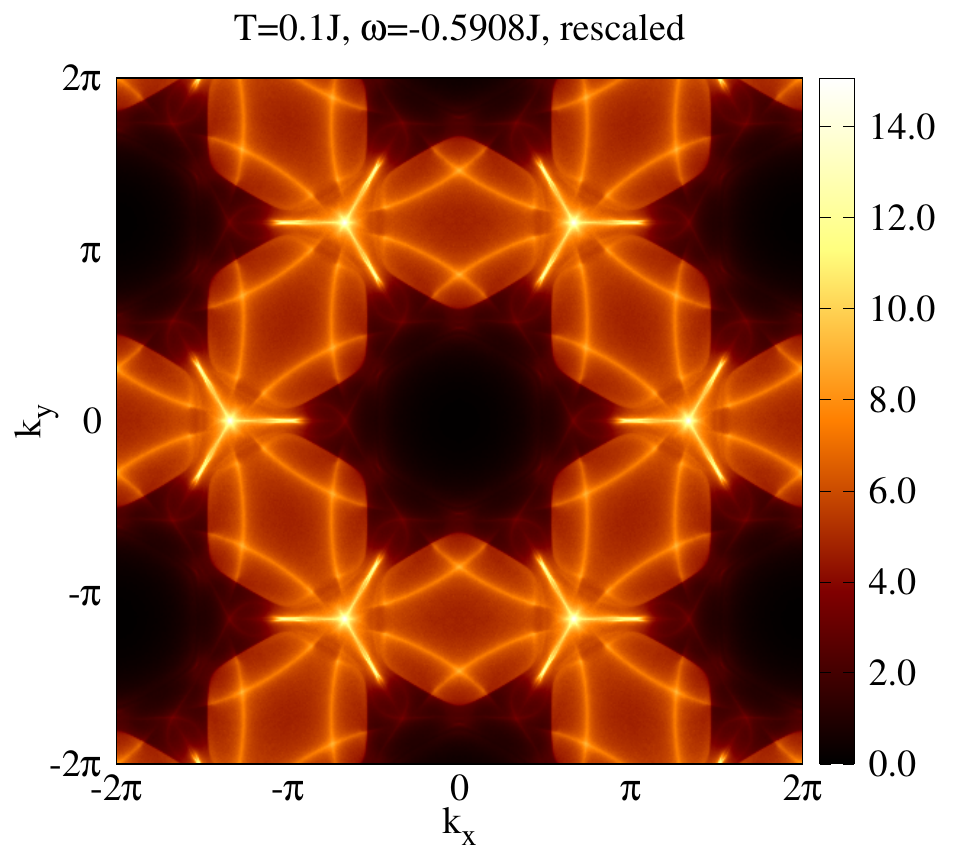}
 \caption{(Color online). $S(\mathbf{k},\omega)$ (top panel) and $\exp(\omega/T)S(\mathbf{k},-\omega)$ (bottom panel) for the $\mathbf{q}=\bm{0}$ Ansatz with $T=0.1J$ and $\omega=0.5908J$. Both results compare very well as per the relation of detailed balance~\eqref{eq:DB}.
 }
 \label{fig:DB}
\end{figure}
As discussed in the main text, at higher temperatures where the spin density is no longer small enough, interactions between spinons cannot be faithfully neglected, and thus SBMFT is no longer reliable. Here we provide SBMFT results for the DSF at high temperatures that we do not expect to be reliably described by SBMFT. 

In Fig.~\ref{fig:DSF_highT}, we show the DSF at $T=0.15J$ for each Ansatz, where now the major weight of the DSF is around $\omega=0J$ and small momenta around $\mathbf{k}=\bm{0}$, with the regions narrowing and becoming visibly distinct compared to the DSF results for $T=0.1J$ shown in Figs.~\ref{fig:DSF_q0} and~\ref{fig:DSF_Sqrt3}. This indicates that the spectrum starts to become more quasi-elastic with increasing temperature. This becomes even clearer when the temperature is raised to $T=0.19J$, where now the DSF shows three distinct thin high-intensity lines, with the weight focused disproportionately at the $\Gamma$ point. This can be understood by looking at the self-consistent parameters as function of temperature in Fig.~\ref{fig:fields}. At high temperatures such as $T=0.19J$, the system has all its spinon bands almost degenerate with eigenvalue $\lambda$, since $\mathcal{A},\mathcal{B}\approx0$. This means that the Bogliubov matrices $V,X\approx\bm{0}_3$, and thus only $U$ and $Y$ are finite. One thus directly sees that this leads to all terms being negligible except for the second in~\eqref{eq:DSF}, which contains only elements of $U$ and $Y$ (cf.~Appendix~\ref{sec:DSFterms}). This term contributes only around $\omega=0J$, because $\epsilon_{\mathbf{q},\uparrow}^m\approx\epsilon_{\mathbf{k}+\mathbf{q},\uparrow}^n\approx\lambda$ at this high temperature. The other terms, though negligible, still lead to small contributions around zero frequency and $\omega=\pm2\lambda$. Thus, we see that with higher temperature, the spectrum is quasi-elastic, meaning that spins are more or less completely noninteracting, which is the expected result in the large-temperature limit of a paramagnet. 

Another interesting point is that at temperatures $T\leq0.1J$, the DSF result along the $\Gamma$-M-K-$\Gamma$ high-symmetry lines looks very distinctive from one Ansatz to the other, while at $T/J=0.15$ and $0.19$ one cannot easily separate the Ans\"atze from their DSF. Thus, the SBMFT Ansatz loses its characteristic features at very high temperatures.

We do not go beyond $T=0.19J$, because at higher temperatures $T\geq0.2J$, the bond mean fields $\mathcal{A}=\mathcal{B}=0$, and this is an indication that SBMFT completely fails to describe such a high-temperature disordered phase where nearest-neighbor correlations are absent.\cite{Arovas1988,Auerbach1994}

\section{SSF results}\label{sec:SSF}
The SSF, which is the integral over frequency space of the DSF, is given by
 
\begin{align}
S(\mathbf{k})=\int_{-\infty}^\infty\d\omega\,S(\mathbf{k},\omega),
\end{align}
and in Fig.~\ref{fig:SSF} we show it for the $\mathbf{q}=\bm{0}$ and $\sqrt{3}\times\sqrt{3}$ Ans\"atze at temperatures $T/J=0.01$ and $0.1$. The SSF shows little change with temperature in terms of its characteristic features, save for a small smoothening effect, thus why we do not show it for intermediate temperature values.

As a sanity check, we have moreover numerically verified that our SSF results satisfy the sum rule\cite{Auerbach1994}

\begin{align}
\frac{1}{N}\sum_\mathbf{k}^\text{B.z.}S(\mathbf{k})=\frac{3}{2}\mathcal{S}(\mathcal{S}+1).
\end{align}
This is also supplemented by a further check, that of detailed balance discussed in Appendix~\ref{sec:DB}.

\section{Detailed balance}\label{sec:DB}
Detailed balance is a relation of the DSF,\cite{DeNardis2016} and is given by

\begin{align}\label{eq:DB}
S(\mathbf{k},\omega)=S(\mathbf{k},-\omega)\text{e}^{\omega/T}.
\end{align}
We numerically check that it is satisfied, and here we provide an example in Fig.~\ref{fig:DB} for the $\mathbf{q}=\bm{0}$ Ansatz at $T=0.1J$ and $\omega=0.5908J$ showing that~\eqref{eq:DB} is indeed satisfied.

\bibliography{fTDSFbiblio}
%--------------------------------------------------------
\end{document}